%% file: mbar.tex
\documentclass[11pt]{article}
\usepackage{epsfig,sint,macros,cite}
\begin{document}
\input title.tex

\input sect1.tex

\input sect2.tex

\input sect3.tex

\input sect4.tex

\input sect5.tex

\input sect6.tex

\input sect7.tex
\begin{appendix}
\input appendix1.tex

\input appendix2.tex

\input appendix3.tex

\end{appendix}
\bibliography{biblist}        
\bibliographystyle{h-elsevier}   
\end{document}

%% file: title.tex
\begin{titlepage}
\begin{flushright}
  DESY-98-154\\
  OUTP-98-68-P
\end{flushright}

\vskip 0.5 cm
\begin{center}
  {\Large\bf Non-perturbative quark mass renormalization\\[0.5ex]
  in quenched lattice QCD\\[0.5ex] }
\end{center}
\vskip 0.5 cm
\vbox{
\centerline{
\epsfxsize=2.5 true cm
\epsfbox{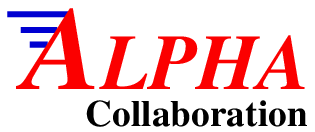}}
}
\vskip 0.5 cm
\begin{center}
{\large Stefano Capitani$^{\scriptscriptstyle a}$, 
     Martin L\"uscher$^{\scriptscriptstyle a}$,
     Rainer Sommer$^{\scriptscriptstyle b}$ and
     Hartmut Wittig$^{\scriptscriptstyle c,}$\footnote{PPARC Advanced Fellow}
\vskip 0.5cm
$^{\scriptstyle a}$
Deutsches Elektronen-Synchrotron, DESY \\
Notkestrasse 85, D-22603 Hamburg, Germany
\vskip 2.5ex
$^{\scriptstyle b}$
DESY-Zeuthen \\
Platanenallee 6, D-15738 Zeuthen, Germany
\vskip 2.5ex
$^{\scriptstyle c}$
Theoretical Physics, University of Oxford \\
1~Keble Road, Oxford OX1~3NP, UK
\vskip 1.0cm
{\bf Abstract}}
\vskip 0.7ex
\end{center}

The renormalization factor relating the bare to the renormalization
group invariant quark masses is accurately calculated in quenched
lattice QCD using a recursive finite-size technique. The result is
presented in the form of a product of a universal factor times another
factor, which depends on the details of the lattice theory but is easy
to compute, since it does not involve any large scale differences. As
a byproduct the $\Lambda$-parameter of the theory is obtained with a
total error of $8\%$.

\vfill

\begin{center}
 October 1998
\end{center}

\eject

\vfill

\eject

\end{titlepage}

%% file: sect1.tex
\section{Introduction}

The masses of the light quarks are not directly accessible to
experiment and have to be determined through a non-perturbative
calculation, taking low-energy hadronic data (such as the kaon masses)
as input. Chiral perturbation theory is able to predict their ratios
with a precision of a few percent \cite{leutwyler:1996}. In order to
obtain estimates of individual quark masses, it thus suffices to
determine a particular linear combination using lattice QCD
\cite{quark:APE1,quark:APE2,quark:APE3,hadr:ukqcd,quark:gupta,quark:gough,impr:qcdsf,impr:qcdsf_lat98,quark:CPPACS,impr:roma2,quark:marti,quark:SESAM}
or QCD sum rules
\cite{quark:bijnens,quark:jamin1,quark:jamin2,quark:narison,quark:chetyrkin,quark:colangelo,quark:prades,quark:yndurain,quark:dosch,quark:lellouch}.

Calculations of quark masses in lattice QCD are in principle
straightforward, but there are several sources of systematic errors
which must be studied carefully (for recent reviews of the subject see
\cite{lat97:gupta,reviews:rajan}). One of the principal uncertainties
arises from the renormalization constant needed to convert from
lattice normalizations to the $\msbar$ scheme of dimensional
regularization. Usually this factor is only known in one-loop
perturbation theory in the bare coupling. The limitations of lattice
perturbation theory are well known, and in order to remove all doubts
about the reliability of quark mass calculations in lattice QCD, it is
evident that a non-perturbative determination of the renormalization
factor is required.

The general problem of non-perturbative renormalization of QCD
addresses the question how the perturbative regime of QCD is related
to the observed hadrons and their interactions. This relation involves
large scale differences, and thus the problem is not easily approached
using numerical simulations, for which only a limited range of lattice
spacings is accessible. 

A strategy how to overcome this technical difficulty has been
described in \cite{impr:lett}, and a comprehensive introduction into
the subject is presented in \cite{reviews:leshouches}. 
An alternative method was introduced in \cite{renorm_mom:paper1}
and is actively being pursued
\cite{quark:marti,renorm_mom:QCDSF,renorm_mom:APEbilin,quark:APE3}. 
In our strategy the key idea is the introduction of an intermediate
renormalization scheme, with a controlled (non-perturbative) relation
to the lattice normalizations, and in which the scale evolution of
masses and couplings can be computed non-perturbatively from low to
very high energies. In the high-energy regime one can continue the
scale evolution using perturbation theory, which allows for the
determination of the renormalization group invariant quark masses~$M$
and the $\Lambda$-parameter. Since the matching of $M$ and $\Lambda$
between different renormalization schemes is exactly computable, it is
evident that all reference to the intermediate scheme is eliminated at
this stage.

Thus, the problem of non-perturbative quark mass renormalization can
be split into two parts. The first is the computation of the scale
dependence of the running masses~$\mbar$ in the intermediate scheme
from low to high energies and its relation to the renormalization
group invariant quark masses~$M$. The ratio $\mbar/M$ is then known at
all energy scales covered by the calculation. The second part is the
matching of the renormalized masses~$\mbar$ with the bare current
quark masses~$m(g_0)$. This amounts to the computation of
$\mbar/m(g_0)$ at a certain value of the energy scale which is chosen
as the matching point. Since the matching can be performed at low
energies, no excessively large lattices are required to contain all
relevant scales. Whereas the second part clearly depends on the
details of the lattice regularization, such as the chosen lattice
action and the bare coupling, the ratio $\mbar/M$ is a universal
quantity.

In this work we have calculated the regularization independent factor
$\mbar/M$ in quenched QCD, using the Schr\"odinger functional (SF) as
our intermediate renormalization scheme. We emphasize that the SF
scheme is chosen entirely for practical purposes, since it allows the
non-perturbative computation of the scale dependence over several
orders of magnitude in a controlled way, using a recursive finite-size
scaling technique and numerical simulations. The numerical results for
the scale dependence of $\mbar$ can be extrapolated reliably to the
continuum limit, so that a truly universal result for $\mbar/M$ is
obtained.

In addition, we have computed the matching between the renormalized
quark masses in the SF scheme and the bare quark masses in $\Oa$
improved lattice QCD~\cite{impr:pap3} for a range of bare couplings at
a fixed scale. Thus, by combining the universal ratio $\mbar/M$ with
the matching factor $\mbar/m(g_0)$ we have obtained the total
renormalization factor $M/m(g_0)$, which relates the bare current
quark masses to the renormalization group invariant quark masses.

We also report on our determination of the $\Lambda$-parameter in
quenched QCD. For this purpose, we have supplemented the numerical
data for the running coupling published in \cite{alpha:SU3}, so that
the scale evolution could be traced over more than two orders of
magnitude.  Preliminary results of our calculations have been presented
in \cite{lat97:martin}.

The rest of the paper is organized as follows. In~\sect{s_runn_PT} we
discuss the scale dependence of the running coupling and quark masses
in perturbation theory. The \SF\ scheme is described in detail
in~\sect{s_SF}. The step scaling functions, which describe the scale
evolution of the running parameters, are briefly discussed
in~\sect{s_ssf}. In \sect{s_runn} we describe the extraction of
$\mbar/M$ and the $\Lambda$-parameter. The matching of these results
to hadronic schemes is presented in~\sect{s_match}, and~\sect{s_concl}
contains our conclusions. In order to avoid distracting the reader
from the main results, we defer all technical details about the
computation of the step scaling functions, the error propagation in
the scale evolution and the calculation of $\mbar/m(g_0)$ to
Appendices~\ref{a_sigmas}, \ref{APX_sigma_fit} and
\ref{a_zp}, respectively.

%% file: sect2.tex
\section{Running coupling and quark masses in perturbation theory
          \label{s_runn_PT}}

\subsection{Mass-independent renormalization schemes}
QCD is a theory with $\nf+1$ free parameters: its coupling constant
$g$ and the masses of the $\nf$ different quark flavours, $\{
m_s,s=1,\ldots,\nf\}$. In the definition of the theory through a
regularization, these are initially taken to be the bare parameters in
the Lagrangian. Upon removal of the regularization, one defines
renormalized parameters, $\gbar,\, \{\mbar_s, s=1,\ldots,\nf\}$ at
some energy scale $\mu$. In the following we assume that the
normalization conditions which are imposed to define $\gbar, \mbar_s$
are independent of the quark masses themselves. Examples for such
mass-independent renormalization schemes are the $\MSbar$ scheme of
dimensional regularization~\cite{ms:thooft,msbar:gen} and the
Schr\"odinger functional (SF) scheme~\cite{mbar:pert,impr:lett}. The
latter will be described in \sect{s_SF}. The renormalized running
parameters are then functions of the renormalization scale $\mu$
alone, and their scale evolution is given by the renormalization group
equations
\bes
   \mu {\partial \gbar \over \partial \mu} &=& \beta(\gbar)\,, 
\label{f_beta_def} \\
   \mu {\partial \mbar_s \over \partial \mu} &=&\tau(\gbar) \mbar_s \, ,
   \quad s=1,\ldots,\nf \,.
\label{f_tau_def}
\ees
The renormalization group functions, $\beta$ and $\tau$, have a
perturbative expansion
\bes
   \beta(\gbar) &\buildrel {\gbar}\rightarrow0\over\sim\,&
 -{\gbar}^3 \left \{ b_0 + b_1 {\gbar}^{2} +  b_2 {\gbar}^{4} + 
   \ldots\right \},  
\label{f_beta_pert} \\
   \tau(\gbar) &\buildrel {\gbar}\rightarrow0\over\sim\,&
 -{\gbar}^2 \left \{ d_0 + d_1 {\gbar}^{2} +  d_2 {\gbar}^{4} + 
   \ldots\right \}\,.
\label{f_tau_pert}
\ees
The universal coefficients $b_0,\,b_1$ and $d_0$ are given by
\bes
  && b_0 = (4\pi)^{-2} \left\{11 - \frac23 \nf \right\}\,, \; 
   b_1 = (4\pi)^{-4} \left\{102 - \frac{38}{3} \nf \right\}\,, \\
  && d_0 = 8/(4\pi)^{2}  \,, 
\ees
whereas the higher order coefficients $b_2,\,b_3,\ldots$ and
$d_1,\,d_2,\ldots$ depend on the chosen renormalization scheme.
Equations~(\ref{f_beta_def},\ref{f_tau_def}) can be solved
straightforwardly. This yields the exact relations between the scale
dependent quantities $\gbar(\mu),\mbar_s(\mu)$ and the renormalization
group invariants,
\bes
  \Lambda &=& \mu\,(b_0\gbar^2)^{-b_1/2b_0^2} \rme^{-1/2b_0\gbar^2} 
  \times
  \exp\left\{-\int_0^{\gbar}\rmd g
  \left[{1\over\beta(g)}+{1\over b_0g^3}-{b_1\over b_0^2 g}\right]\right\}, 
\label{f_Lambda_def} \\
  \noalign{\vskip3ex}
  M_s &=& \mbar_s\,(2 b_0\gbar^2)^{-d_0/2b_0} 
  \times
  \exp\left\{-\int_0^{\gbar}\rmd g 
  \left[{\tau(g)\over\beta(g)}-{d_0\over b_0g}\right]\right\}.
\label{f_M_def}
\ees
Unlike the case of the $\Lambda$-parameter, there is no universally
accepted normalization convention for the renormalization group
invariant quark masses~$M_s$. \Eq{f_M_def} complies with the
conventions used by Gasser and Leutwyler
\cite{GasLeut}. 

In contrast to $\gbar$ and $\mbar_s$, the renormalization group
invariants $\Lambda$ and $M_s$ do not depend on the scale; from a
technical point of view they are the integration constants of the
differential equations (\ref{f_beta_def},\ref{f_tau_def}). From
\eq{f_M_def} and the scale independence of $M_s$, we read off
immediately that ratios of running quark masses,
$\mbar_s/\mbar_{s^\prime}$, are scale independent in mass-independent
renormalization schemes, and are equal to $M_s/M_{s^\prime}$.

The renormalization group invariant quark masses are easily shown to
be independent of the renormalization scheme, while the
$\Lambda$-parameter changes from scheme to scheme by factors that can
be calculated exactly. It is hence natural to take $\Lambda$ and the
renormalization group invariant quark masses $M_s$ as the fundamental
parameters of QCD.

Furthermore one concludes that in order to extract the
fundamental parameters of QCD, one may choose a scheme which
is particularly suited for the computation of $\Lambda$ and~$M_s$.

\begin{figure}[tb]
\hspace{0cm}
\vspace{-4.2cm}

\centerline{
\psfig{file=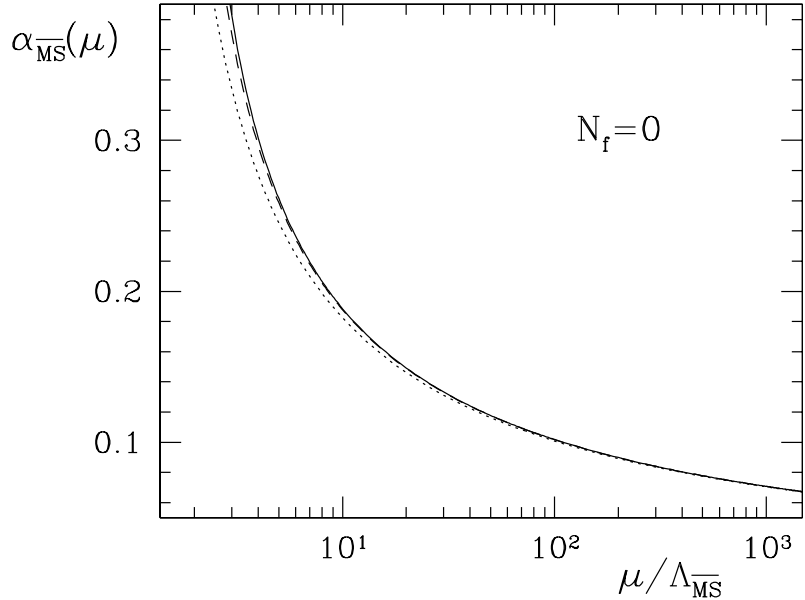,width=14cm}
}
\vspace{-1.2cm}
\caption{\footnotesize
Running coupling $\alpha_{\msbar}=\gbar^2_{\msbar}/4\pi$ 
in quenched QCD.
From the dotted to the full curve, the perturbative accuracy increases
from two to four loops.
\label{f_alphaMS}}
\end{figure}

\begin{figure}[tb]
\hspace{0cm}
\vspace{-4.2cm}

\centerline{
\psfig{file=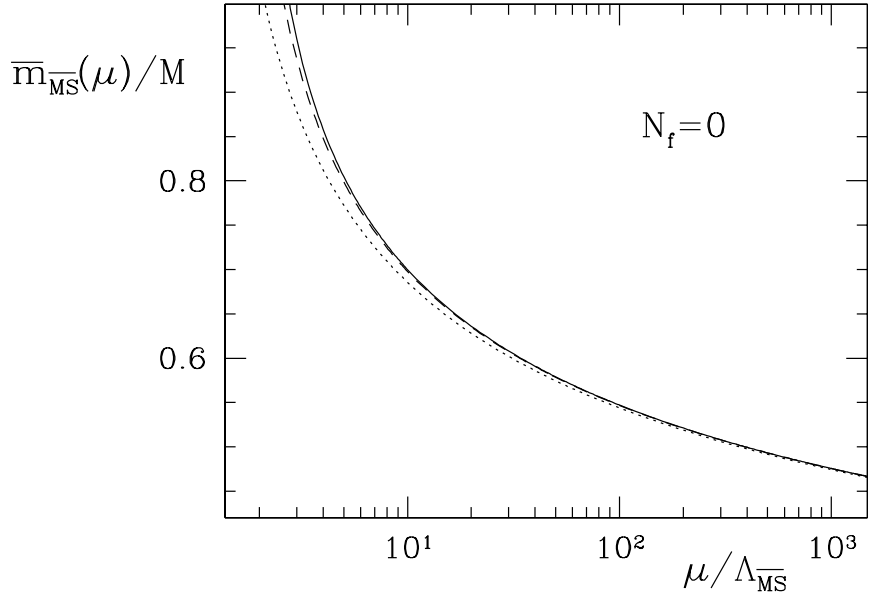,width=14cm}
}
\vspace{-1.2cm}
\caption{\footnotesize
Running quark mass in quenched QCD. The flavour index has 
been omitted, since the entire graph is independent of which quark
flavour is considered.
\label{f_mbarMS}}
\end{figure}
%
\subsection{Illustration: the $\MSbar$ scheme}
In the $\MSbar$ scheme the perturbative expansions
eqs.~(\ref{f_beta_pert},\ref{f_tau_pert}) are known to four loops
\cite{MS:2loop1,MS:2loop2,MS:2loop3,MS:2loop4,MS:3loop1,MS:3loop2,MS:3loop3,MS:3loop4,MS:4loop1,MS:4loop2,MS:4loop3}.
Here we use them to plot the evolution of the running coupling and
quark masses in this scheme for $\nf=0$ flavours. The dotted, dashed
and full curves in Figs.~\ref{f_alphaMS} and \ref{f_mbarMS} have been
obtained from eqs.~(\ref{f_Lambda_def}) and~(\ref{f_M_def}) by
substituting the 2-, 3- and 4-loop expressions for the $\beta$- and
$\tau$-functions and computing the integrals in the exponents
numerically.

Conventionally the running coupling and quark masses are quoted in the
$\MSbar$ scheme at a certain value of the normalization
mass~$\mu$. Figures~\ref{f_alphaMS} and~\ref{f_mbarMS} illustrate
that, once $\Lambda$ and~$M$ are known, the running parameters in the
$\MSbar$ scheme are uniquely determined to any given order of
perturbation theory.

As we shall see in~\sect{s_match} the value of $\Lambda_{\msbar}$ in
quenched QCD is about $240\,\MeV$. From Figs.~\ref{f_alphaMS}
and~\ref{f_mbarMS} one infers that perturbation theory appears to
``converge'' at energies as low as $1\,\GeV$. However, since the
$\MSbar$ scheme is only defined to any finite order of perturbation
theory, it is impossible to make a solid statement about the total
uncertainty in the running coupling and quark masses at a certain
reference scale. In other words, the running coupling and quark masses
in the $\MSbar$ scheme are only meaningful to a given order of
perturbation theory.

%% file: sect3.tex
\section{The Schr\"odinger functional scheme\label{s_SF}}

In this section we describe the intermediate renormalization scheme,
which we have used to relate hadronic observables to the
renormalization group invariant parameters. The chosen scheme, defined
using the \SF, has been the subject of a series of publications
\cite{SF:LNWW,SF:stefan1,SF:stefan2,alpha:SU3,pert:1loop,mbar:pert}.
For clarity we repeat the basic concepts of the \SF\ but shall
otherwise be rather brief and refer the reader to the literature 
for details. A pedagogical introduction can be found in
\cite{reviews:leshouches}.

\subsection{General definitions}

As already mentioned in the introduction, the main difficulty of
relating the observed hadron spectrum to the perturbative regime of
QCD in a controlled way is the large scale difference.
Non-perturbative renormalization such as the approach introduced
in~\cite{renorm_mom:paper1} relies on all relevant physical scales
being accommodated on a single lattice. This results in a rather
narrow energy range which can be explored.

It has been demonstrated \cite{alpha:sigma} that this problem can be
overcome by simulating a sequence of lattices with decreasing lattice
spacing. Any single lattice only covers a limited range of distances,
but through the use of a finite-volume renormalization scheme it is
possible to match subsequent lattices. With only a few steps of this
procedure one can easily cover a much larger energy range than
otherwise possible.

The \SF\ is a particular finite-volume renormalization scheme. It is
based on the formulation of QCD in a finite space-time volume of
size $T\times L^3$, with periodic spatial boundary conditions and
inhomogeneous Dirichlet boundary conditions at time $x_0=0$ and
$x_0=T$ \cite{SF:LNWW,SF:stefan1}. This is realized by requiring the
spatial components of the gauge field at the boundary to be equal to
some prescribed constant abelian fields $C$ and $C^\prime$. This
choice of boundary condition introduces a frequency gap on the quark
and gluon fields, so that simulations for vanishing quark mass can be
performed. 

We take over the exact form of the action and the boundary conditions
from Sect.~2.1 of ref.~\cite{pert:1loop}; any notation not explained
here is taken over from that reference. The definition of the scheme
requires that the ratio $T/L$ is kept fixed.  For reasons that will be
explained below we have set $T=L$ throughout this work. Within this
set-up, renormalization conditions are then specified at scale
$\mu=1/L$ and vanishing quark mass. Thus, the \SF\ is a
mass-independent renormalization scheme, in which the running coupling
and masses scale with the box size $L$.

\subsection{Running coupling and masses}

Our next task is to specify exactly the definition of the running
coupling and quark masses in the SF scheme, which we have used in the
numerical simulations. In particular, we will exploit the freedom in
choosing the boundary fields $C$ and $C^\prime$, and the angle
$\theta$ which appears in the spatial boundary conditions of quark
fields. From a practical point of view, these have to be chosen with
care in order to ensure that the scale evolution of masses and
couplings can be determined with high accuracy in the continuum limit
and that contact with the perturbative evolution at high energies can
easily be made.

The ``optimal'' definitions of $\gbar$ and $\mbar$
\cite{SF:LNWW,alpha:SU3,pert:1loop,mbar:pert} are obtained by
considering the following criteria:

\begin{itemize}
  \item 
  Choose observables with small variance in Monte Carlo simulations to
  ensure good statistical precision of the results;
  
  \item 
  Avoid parameter values which introduce ``accidentally'' large
  coefficients in the perturbative expansion of e.g. the $\beta$- and
  $\tau$-functions;

 \item
  Minimize the contamination of coupling and quark mass by lattice
  artefacts.
\end{itemize}

The renormalized coupling at length scale $L$ is given in terms of an
infinitesimal variation of the boundary conditions for the gauge
fields, exactly as defined in refs.~\cite{alpha:SU3,pert:1loop}, where
all details can be found. It is denoted by $\gbar(L)$, whereas in
accordance with standard notation in the $\msbar$ scheme, we choose
the corresponding energy scale as argument for $\alpha$, viz.
\bes
 \alpha(\mu) = {\gbar^2(L) \over 4 \pi}\,, \;\; \mu=1/L\,.
\ees

We now describe the definition of the running quark mass $\mbar(\mu)$
in more detail and introduce it formally in the framework of continuum
QCD. Its definition is easily made rigorous in the lattice
regularization (see \App{a_sigmap} and \cite{mbar:pert}). The starting
point is the PCAC relation,
\bes
  \partial_{\mu} (\ar)_{\mu} = (\mbar_{s} + \mbar_{s'}) \pr  \,,
 \label{e_PCAC}
\ees
between the renormalized axial current,
\bes
  \left(\ar\right)_{\mu}(x) = 
                 \za \,\psibar_s(x) \gamma_{\mu} \gamma_5 \psi_{s'}(x),
\ees
and the pseudoscalar density,
\bes
  \pr(x) = 
                 \zp \,\psibar_s(x) \gamma_5 \psi_{s'}(x)\,.
\ees
Once \eq{e_PCAC} is chosen to define the quark masses, their
normalization is given in terms of the normalization of $\ar$ and
$\pr$. The axial current is normalized naturally through current
algebra relations
\cite{Boch,MaiMarti86,Heatlie91,MarSachVla91,MarSachSalVlad92,MarPetSachVlad93,PacPetTagVlad94,UKQCD_za,impr:pap4}
and does not acquire a scale dependence through
renormalization. Therefore, both scheme- and scale-dependence are
contained in the definition of $\pr$.

In order to formulate a normalization condition for the pseudoscalar
density, we set~\cite{mbar:pert},
\bes
  T=L,\quad C=C'=0\,, \quad \theta = 1/2 \,.
\ees
This choice is motivated by the criteria listed above, and a more
detailed explanation is given in ref.~\cite{mbar:pert}. We then start
from the correlation functions
\bes
  \fp(x_0) & = & -\frac{1}{3}\int \rmd^3\vecy\,\rmd^3\vecz\,\langle\,
         \psibar(x)\gamma_5\frac{1}{2}\tau^a\psi(x) \,
         \zetabar(\vecy)\gamma_5\frac{1}{2}\tau^a\zeta(\vecz) \,
         \rangle \label{e_fp} \\
    f_1 & = & -{1\over{3L^6}} \int \rmd^3\vecu\,\rmd^3\vecv
                            \,\rmd^3\vecy\,\rmd^3\vecz \,
        \langle\,
         \zetabarprime(\vecu)\gamma_5\frac{1}{2}\tau^a\zetaprime(\vecv)\,
         \zetabar(\vecy)\gamma_5\frac{1}{2}\tau^a\zeta(\vecz)
         \,\rangle \, ,\label{e_f1} 
\ees
involving the boundary quark fields, $\zeta,\ldots,\zetabarprime$
\cite{impr:pap1}. Here and in the following, the Pauli-matrices,
$\tau^a$, are understood to act on the first two flavour components of
the quark fields. As the renormalization is flavour-blind in a
mass-independent renormalization scheme, this is sufficient to define
the renormalization constants. Furthermore we introduce bare current
quark masses $m_s$ via
\bes
  \partial_{\mu}\, \left[\psibar_s(x) \gamma_{\mu} \gamma_5
  \psi_{s'}(x)\right] &=& 
  (m_s+m_{s'}) \, \psibar_s(x) \gamma_5 \psi_{s'}(x) \, ,
\label{e_mcont}
\ees
and define the renormalization constant
\bes
 \zp(L) &=& {\sqrt{3 f_1} \over \fp(L/2)} \, , \quad
                  {\rm at}\; m_s=0, \,\quad s=1,\ldots,\nf \,.
 \label{e_zp}
\ees
Here, the numerator $\sqrt{f_1}$ is introduced to cancel the
multiplicative renormalization of the boundary quark fields
$\zeta$ and $\zetabar$, which appear in the definition of $\fp$. Our
convention \eq{e_zp} ensures that $\zp = 1$ at tree level of
perturbation theory.

Explicitly the definition of the running quark masses in the SF scheme
now reads
\bes
  \mbar(\mu)_s &=& {\za \over \zp(L)} m_s\,, \quad \mu=1/L\,. \label{e_mbar}
\ees

The above expressions are easily given a precise meaning in the
lattice regularization, and we refer to the formulae listed in
\App{a_sigmap_def}. We emphasize, however, that the SF-scheme is not
linked to a particular regularization.  Note that $\zp$ is defined in
terms of correlation functions at finite physical distances for which
on-shell $\Oa$ improvement may be
applied~\cite{impr:onshell,impr:pap1} to reduce lattice
artefacts. Although this detail is irrelevant for the definition of
the scheme itself, it will be important in its numerical
implementation.

From now on we will drop the flavour index~$s$ on all quark masses,
except when we refer to a particular flavour.

Finally we note that the normalization constant $\za$ of the axial
current is known non-perturbatively in $\Oa$ improved quenched lattice
QCD \cite{impr:pap4}. Hence, in order to determine the 
complete renormalization factor in \eq{e_mbar}, it suffices to compute 
the renormalization constant $\zp$.

\subsection{Perturbation theory}

The SF scheme has been studied perturbatively in quite some detail.
In the rest of this section we list the results which are relevant for
the present computation, setting $\nf=0$. All quantities where we do
not indicate the scheme explicitly refer to the SF scheme.  The
$\Lambda$-parameter is translated to the $\msbar$ scheme via
\cite{alpha:SU3}
\bes
  \Lambda  = 0.48811(1) \Lambda_{\msbar} \,, \label{e_Lambdaratio}
\ees
and the three-loop coefficient of the $\beta$-function was determined
as~\cite{pert:2loop}
\bes
  b_2 = 0.483(9)/(4\pi)^3 \,. \label{e_b2}
\ees
In addition, the two-loop anomalous dimension of the quark mass,
\bes
  d_1 = 0.217(1)/(4\pi)^2 \,, \label{e_d1}
\ees
was computed as part of this project \cite{mbar:pert}. Given the
perturbative expansion coefficients of $\beta$ and $\tau$ in the
$\msbar$ scheme, these equations fix the one-loop relation of $\mbar$
to $\mbar_{\msbar}$ and the two-loop relation of $\alpha$ to
$\alphaMSbar$. Explicit formulae may be found
in~\cite{mbar:pert,pert:2loop}.

For our purposes, the coefficients $b_2$ and $d_1$ serve primarily to
continue the scale evolution of the running coupling and quark mass in
the SF scheme to infinite energy using perturbation theory. This
allows for the extraction of $\Lambda$ and $M$, which can be easily
converted into their counterparts in the $\msbar$ scheme through
\eq{e_Lambdaratio} and the fact that $M=M_{\msbar}$.

%% file: sect4.tex
\section{Step scaling functions \label{s_ssf}}

In the SF scheme a change of the renormalization scale amounts to a change
of the box size $L$ at fixed bare parameters. By considering a
sequence of pairs of volumes with sizes $L$ and $2L$, one can thus study the
evolution of the running coupling and quark masses under repeated
changes of the scale by a factor $2$. Effectively one
constructs a non-perturbative renormalization group in this way.

The evolution from size $L$ to
$2L$ of the running coupling and the normalization 
of the pseudoscalar density is described by the step scaling functions 
$\sigma(u)$ and $\sigmap(u)$ according to
\bes 
  \gbar^2(2L) &=& \sigma(u), \qquad u\equiv\gbar^2(L),  \\
  \noalign{\vskip1ex}
  \zp(2L)     &=& \sigmap(u)\,\zp(L).
\ees
For a given lattice resolution $a/L$ both functions can
be computed non-perturbatively through numerical simulations of the
Schr\"odinger functional. It is important to realize that the box size $L$
can be as small as we like in these calculations.
The only restriction is that $L$ and the low-energy
physical scales in the theory should be significantly larger than the
lattice spacing to avoid uncontrolled cutoff effects.

\input sigma_tab

The step scaling function associated with the coupling
has first been computed in ref.~\cite{alpha:SU3}.
We have extended these calculations
so that $\sigma(u)$ is now known precisely 
for altogether $11$ values of $u$ (see \tab{TAB_sigma}).
Moreover we have obtained $\sigmap(u)$ at the same
couplings and, in  addition, at $u=2.77$ and $u=3.48$. 
The technical details of our computations are 
summarized in Appendix~\ref{a_sigmas}. Here we only mention that
the lattice results for the step scaling functions 
depend on the lattice spacing and have to be extrapolated
to the continuum limit. The extrapolation is required but not critical,
since in all cases considered the observed 
lattice effects are small, particularly after including the 
relevant O($a$) correction terms.

%% file: sigma_tab.tex
\begin{table}[tb]
\centering
\begin{tabular}{l l l l l}
\hline \\[-1.5ex]
\multicolumn{1}{c}{$u$}
&\hspace{2em}
&\multicolumn{1}{c}{$\sigma(u)$} 
&\hspace{2em}
&\multicolumn{1}{c}{$\sigmap(u)$} \\[1.0ex]
\hline \\[-1.5ex]
0.8873  &&    0.981(9)     &&    0.9683(21)  \\ 
0.9944  &&    1.110(11)    &&    0.9672(23)  \\
1.0989  &&    1.252(11)    &&    0.9622(24)  \\
1.2430  &&    1.416(16)    &&    0.9579(29)  \\
1.3293  &&    1.528(20)    &&    0.9470(28)  \\
1.4300  &&    1.703(24)    &&    0.9407(30)  \\
1.5553  &&    1.865(23)    &&    0.9382(33)  \\
1.6950  &&    2.095(25)    &&    0.9297(32)  \\
1.8811  &&    2.412(32)    &&    0.9284(36)  \\
2.1000  &&    2.771(41)    &&    0.9168(37)  \\
2.4484  &&    3.464(40)    &&    0.8942(38)  \\
2.7700  &&                 &&    0.8781(42)  \\
3.4800  &&                 &&    0.8451(55)  
  \\[1.0ex]
\hline
\end{tabular}
\caption[TAB_sigma]{\footnotesize 
Simulation results for the step scaling functions $\sigma(u)$ and 
$\sigmap(u)$
\label{TAB_sigma}
}
\end{table}

%% file: sect5.tex
\section{Running coupling and quark mass in the SF scheme\label{s_runn}}

Having computed the step scaling functions, we can now 
work out the evolution of the coupling and the quark masses
over roughly two decades of the energy scale.
It should be stressed from the beginning that 
all results obtained in this section refer to the continuum theory.
The basic idea is to start from some initial values at low energies
and to step up the energy scale by factors of $2$ by repeated application of 
the step scaling functions. In terms of the box size $L$ this
leads us from some large size $\Lmax$ to smaller sizes
$2^{-k}\Lmax$. Eventually the perturbative
regime is reached and perturbation theory may then be used
to extract the $\Lambda$-parameter and the 
renormalization group invariant quark masses.

\begin{figure}[tb]
\hspace{0cm}
\vspace{-4.2cm}

\centerline{
\psfig{file=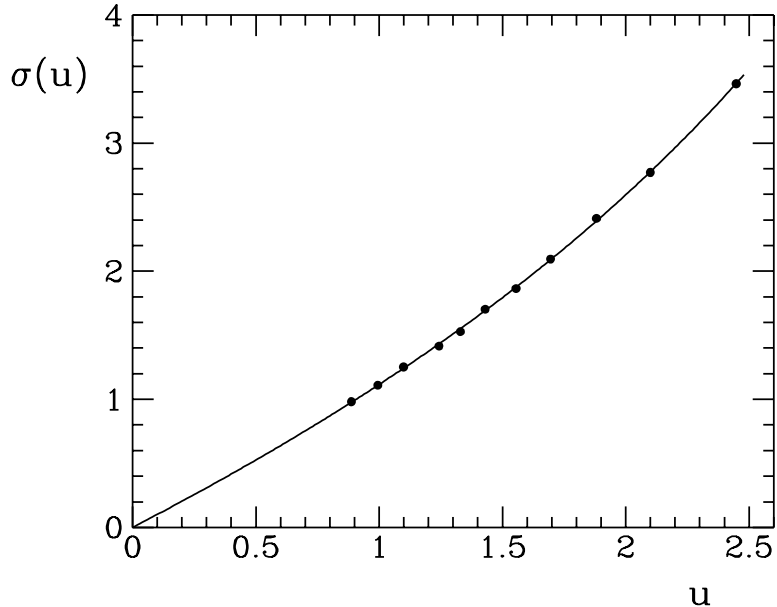,width=14cm}
}
\vspace{-1.2cm}
\caption{\footnotesize
Polynomial fit of the data for the step scaling function $\sigma(u)$.
The errors on the data are about equal to the symbol size.
\label{FIG_sigma_fit}}
\end{figure}

The largest value of $\gbar^2$ which can be 
reached with the available data for the step scaling function
$\sigma(u)$ is $3.48$. We thus define the scale $\Lmax$ through
\be
   \gbar^2(L)=3.48\quad\hbox{at}\quad L=\Lmax.
   \label{Lmax_def}
\ee
The sequence of couplings\footnote{Note that, contrary to the
convention used in \cite{lat97:martin}, the box size~$L$ decreases for
increasing $k$.}
\be
   u_k=\gbar^2(2^{-k}\Lmax),\qquad k=0,1,2,\ldots,
\ee
can then be calculated by solving the recursion
\be
   u_0=3.48,\qquad \sigma(u_{k+1})=u_k.
   \label{u_recursion}
\ee
A technical problem here is that $\sigma(u)$ is only known at certain
values of the coupling and only to a finite numerical precision. 
This difficulty can be resolved by fitting the data
with a polynomial, as shown in \fig{FIG_sigma_fit},
and using the fit function in the recursion (\ref{u_recursion}).
The analysis presented in \App{APX_sigma_fit}
shows that the systematic uncertainty which is incurred by 
this procedure is negligible compared to the statistical
errors, provided one stays in the range of couplings
covered by the data. 

\begin{figure}[tb]
\hspace{0cm}
\vspace{-4.2cm}

\centerline{
\psfig{file=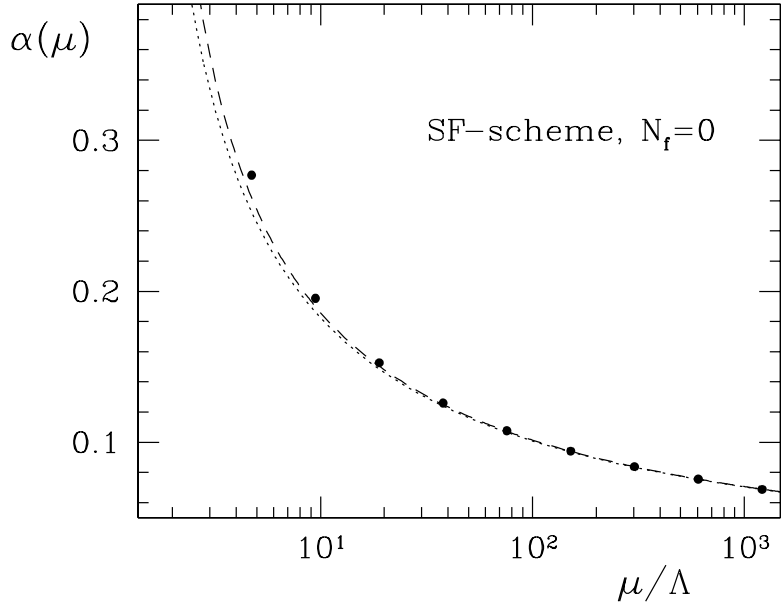,width=14cm}
}
\vspace{-1.2cm}
\caption{\footnotesize
Comparison of the numerically computed values of the running coupling 
in the SF scheme with perturbation theory.
The dotted and dashed curves are 
obtained from \eq{f_Lambda_def} using the $2$- and $3$-loop expressions
for the $\beta$-function. The errors on the data are smaller
than the symbol size.
\label{FIG_alpha_run}}
\end{figure}

After six steps the recursion yields
\be
  \gbar^2(L)=1.053(12)\quad\hbox{at}\quad L=2^{-6}\Lmax
  \label{gbar_high}
\ee  
and two more iterations bring us down to a value of $0.865(11)$.
As will become clear below, these couplings are deep in the 
high-energy regime where the scale evolution is accurately
described by perturbation theory. 
The $\Lambda$-parameter may thus be calculated 
by setting $\mu=1/L$ in {\eq{f_Lambda_def} and inserting 
the coupling quoted above. To evaluate the integral in this formula,
one may safely use the 3-loop expression for the $\beta$-function
since the coupling is small in the whole integration range
(if an approximately geometric progression of the coefficients
$b_k$ is assumed, the error from the neglected higher-order terms 
is an order of magnitude smaller than the statistical error).
As result one obtains
\be
  \Lambda=0.211(16)/\Lmax
  \label{Lambda_Lmax}
\ee
and from \fig{FIG_alpha_run} one can now see that the numerically
determined values of the coupling are indeed closely matched
by the perturbative evolution in the range where the $\Lambda$-parameter
has been extracted. In this plot the error in \eq{Lambda_Lmax}
has not been taken into account since it amounts to an overall
scale shift which is of no relevance for the comparison 
of the scale evolution of the coupling with perturbation theory.

\begin{figure}[tb]
\hspace{0cm}
\vspace{-4.2cm}

\centerline{
\psfig{file=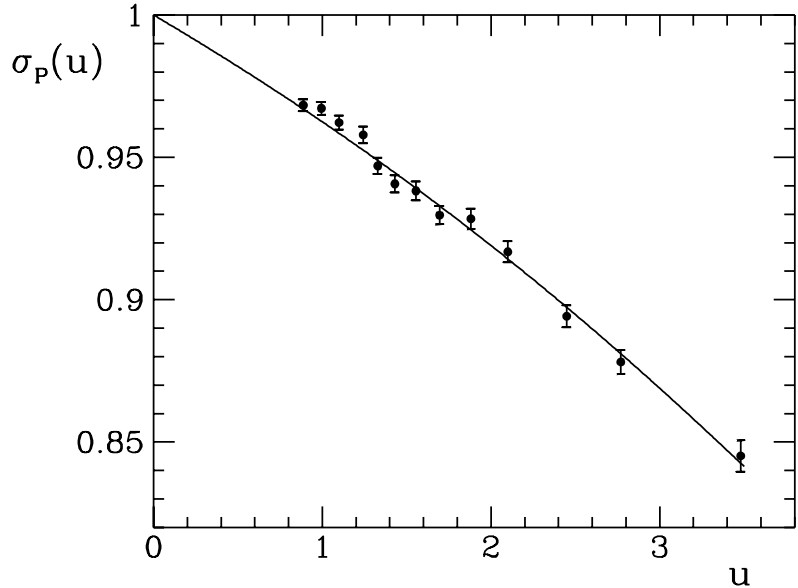,width=14cm}
}
\vspace{-1.2cm}
\caption{\footnotesize
Polynomial fit of the data for the step scaling function $\sigmap(u)$.
\label{FIG_sigmap_fit}}
\end{figure}

We now proceed to discuss the scale dependence 
of the renormalized pseudoscalar density and the running quark mass.
In this case the data for the step scaling functions listed in 
\tab{TAB_sigma} allow us to trace the scale evolution up to $L=2\Lmax$.
If we define
\be
  v_k=\zp(2^{-k+1}\Lmax)/\zp(2\Lmax), 
  \qquad k=0,1,2,\ldots,
\ee
the recursion to be solved is
\be
  v_0=1,\qquad
  v_{k+1}=v_k/\sigmap(u_k),
  \label{v_recursion}
\ee
where $u_k$, $k\geq0$, is the sequence of couplings introduced above. 
Using the fit for $\sigmap(u)$ 
displayed in \fig{FIG_sigmap_fit} it is straightforward
to perform this calculation and after seven steps 
one finds
\be
 \zp(L)/\zp(2\Lmax)=1.759(15)
 \quad\hbox{at}\quad L=2^{-6}\Lmax
 \label{zp_high}
\ee
(see \App{APX_sigma_fit} for further details).
At this point the coupling is so small that 
it is safe to insert the two- and three-loop expressions 
for the $\tau$- and $\beta$-function in \eq{f_M_def}.
Taking \eq{zp_high} into account the result
\be
 M/\mbar(\mu)=1.157(12)
 \quad\hbox{at}\quad \mu=(2\Lmax)^{-1}
 \label{M_over_mbar}
\ee
is then obtained. Similar to the coupling,
the scale evolution of the running quark masses in the SF scheme
is accurately reproduced by perturbation theory 
down to surprisingly low energies
(see \fig{FIG_mbar_run}).

\begin{figure}[tb]
\hspace{0cm}
\vspace{-4.2cm}

\centerline{
\psfig{file=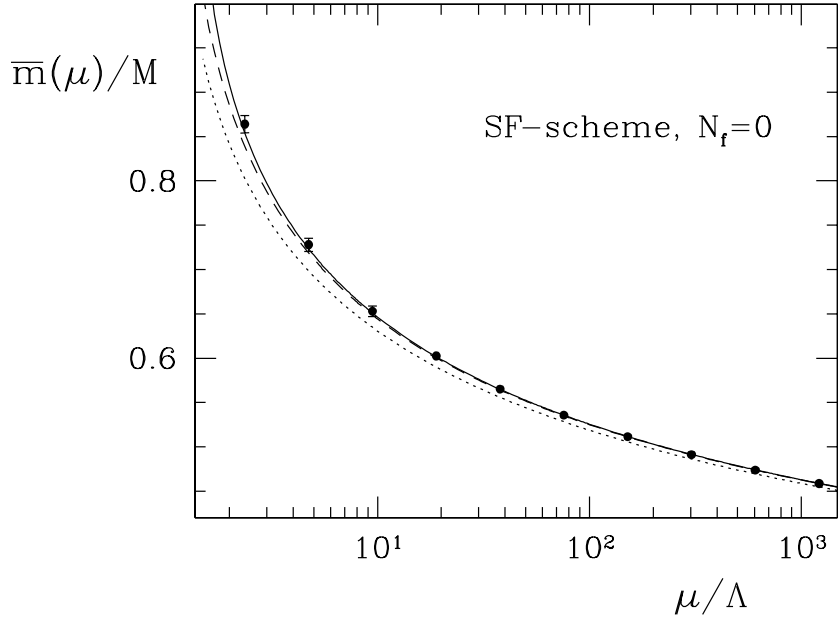,width=14cm}
}
\vspace{-1.2cm}
\caption{\footnotesize
Comparison of the numerically computed values of the running quark mass
in the SF scheme with perturbation theory.
The dotted, dashed and solid curves are 
obtained from eqs.~(\ref{f_Lambda_def}) and
(\ref{f_M_def}) using the $1/2$-, $2/2$- 
and $2/3$-loop expressions
for the $\tau$- and $\beta$-functions respectively.
\label{FIG_mbar_run}}
\end{figure}

%% file: sect6.tex
\section{Matching to hadronic observables \label{s_match}}

At this point the $\Lambda$-parameter and the renormalization group
invariant quark masses~$M$ are known in terms of the reference scale
$\Lmax$ and the running quark masses $\mbar(\mu)$ at
$\mu=(2\Lmax)^{-1}$. To complete the calculation we now need to relate
$\Lmax$ and $\mbar(\mu)$ to a hadronic scheme, where the parameters of
the theory are fixed by requiring a set of low-energy hadronic
quantities to assume their physical values. It should again be
emphasized that this step does not involve any large scale
differences. In particular, the calculations can be carried out using
lattices which cover all relevant scales, as in a conventional hadron
mass computation.

\subsection{Computation of $\Lambda_{\msbar}$}

We now describe how to convert the result for the $\Lambda$-parameter
in~\eq{Lambda_Lmax} into $\Lambda_{\msbar}$ expressed in terms of a
hadronic scale. A convenient hadronic reference scale is the radius
$r_0$ which has been introduced in ref.~\cite{pot:r0}. It is
defined through the force between static quarks and is thus a purely
gluonic quantity in the quenched approximation, similarly to
$\gbar^2(L)$ and $\Lmax$. Relating $\Lmax$ to $\rnod$ is
straightforward, but requires data for $\rnod/a$ with good precision
for a range of $g_0$, in order to obtain the continuum limit
\cite{pot:r0_SU3}. As part of our project, the ratio $\Lmax/\rnod$ was
computed \cite{pot:r0_SU3},
\bes
  \Lmax/\rnod = 0.718(16) \, . \label{e_lmaxrnod}
\ees
We can now remove all reference to the intermediate renormalization
scheme and quote the $\Lambda$-parameter in units of $r_0$. By
combining eqs.~(\ref{e_lmaxrnod}), (\ref{Lambda_Lmax}) and
(\ref{e_Lambdaratio}) we obtain
\bes
  \Lambda_{\msbar}^{(0)} = 0.602(48)/r_0\, ,
  \label{e_lambdaMSr0}
\ees  
where the superscript $^{(0)}$ reminds us that this quantity has been
determined for $\nf=0$, i.e. in the quenched theory. For illustration,
physical units can be introduced by setting $\rnod=0.5\,\fm$, which
gives
\footnote{The value of $\Lambda_{\msbar}^{(0)}$ differs from the result
quoted in \protect\cite{lat97:martin,reviews:leshouches}, because
$\Lmax/r_0$ has been recalculated in ref.~\protect\cite{pot:r0_SU3}.}
\bes
  \Lambda_{\msbar}^{(0)} = 238(19)\,\MeV.
\ees
However, as is well known, the conversion to physical units is
ambiguous in the quenched approximation. The solid result is
\eq{e_lambdaMSr0}.

\subsection{Computation of the total renormalization factor}

Next we describe the calculation of the renormalization factor
$\zM(g_0)$ which relates the bare current quark masses~$m$ to the
renormalization group invariant quark masses~$M$. The renormalized
current quark masses have been defined in \eq{e_mcont}, and their
definition in the $\Oa$ improved lattice theory is given in
\eq{EQ_mPCAC}. By including the appropriate renormalization factors
for the axial current and the pseudoscalar density we can now write
down the relation between~$M$ and~$m$ in $\Oa$ improved lattice QCD,
viz.
\bes
  M  =  {M\over{\mbar(\mu)}}\,
  {{\za(g_0)(1+{\ba}a\mq)}\over{\zp(g_0,L)(1+{\bp}a\mq)}}\, m
  +\Oasq,\quad \mu=1/L.
\ees
Here, the subtracted mass $\mq$ is defined by $\mq=m_0-m_{\rm c}$,
where $m_{\rm c}$ is the critical value of the bare quark
mass~$m_0$ (c.f. ref.~\cite{impr:pap1}).

The improvement coefficients $\ba$ and $\bp$ are defined
in~\cite{impr:pap1}. A strategy how to compute them non-perturbatively
has been described recently in~\cite{impr:rajan_etal}, but no results
for our choice of action have been reported so far. However, the
difference $\ba-\bp$, which is relevant in the above relation has been
shown to be small~\cite{impr:pap5,impr:roma2_1}. Therefore it appears
safe to neglect $\ba$ and $\bp$ altogether, provided that one is
interested in the light quark masses only.

After these considerations we define
\bes
   \zM(g_0) = {M\over{\mbar(\mu)}}\,{{\za(g_0)}\over{\zp(g_0,L)}},
   \quad\mu=1/L,
\label{EQ_ZMdef}
\ees
such that
\bes
  M = \zM(g_0)\,m(g_0) + \Oasq.
\ees
Thus the total renormalization $\zM$ consists of the regularization
independent part $M/\mbar$ and the ratio $\za/\zp$, which depends on
the details of the lattice formulation. In \eq{EQ_ZMdef} all reference
to the SF scheme has disappeared, so that $\zM$ depends only on the
lattice regularization. Furthermore $\zM$, contrary to $\zp$, is
independent of the matching point.

As described in \sect{s_runn} the ratio $M/\mbar$ is obtained for
scales $\mu$ down to a minimum value of $\mu=(2\Lmax)^{-1}$. It is
therefore natural to choose
\bes
  L=1.436\,r_0
\ees
as the value of the matching scale, which is twice the central value
of $\Lmax$ in \eq{e_lmaxrnod}. We obtain
\bes
  M/\mbar(\mu)  = 1.157(15)\quad \mbox{at}
   \quad \mu=(1.436\,\rnod)^{-1}\,.
\label{e_R1}
\ees
The error in \eq{e_R1} contains a small contribution from the error in
\eq{e_lmaxrnod}, which was estimated using the perturbative scale
dependence of $\mbar$.

\begin{figure}[tb]
\hspace{0cm}
\vspace{-6.0cm}
 
\centerline{
\psfig{file=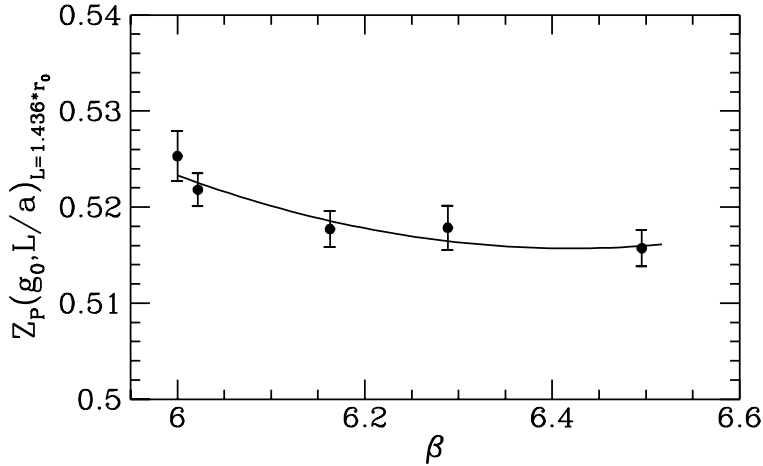,width=14cm}
}
\vspace{-1.0cm}
\caption{\footnotesize
Numerical results for $\zp(g_0,L/a)_{L=1.436\,r_0}$. The solid line
represents the fit \protect\eq{e_zpparam}.
\label{f_zpparam}
}
\end{figure}

The next step in the determination of $\zM$ is the calculation of
$\za(g_0)/\zp(g_0,L)$ at the matching point $L=1.436\,r_0$ for a range
of bare couplings. Results for the normalization constant of the axial
current, $\za$, have been presented in~\cite{impr:pap4}. Eq.~(6.11) in
ref.~\cite{impr:pap4} describes $\za(g_0)$ with an accuracy of 1\,\%
for $g_0\leq1$.

We have now also computed $\zp$ in the range $6.0 \leq \beta \leq
6.5\,,\;\beta=6/g_0^2$. The details of the calculation are explained
in \App{a_zp}, and results listed in \tab{TAB_zp} and in
\eq{e_zp_beta6}. They are represented by a polynomial fit
\bes
& &\zp(g_0,L/a)_{L=1.436\,r_0} = 0.5233-0.0362\,(\beta-6)
                                +0.0430\,(\beta-6)^2, \nonumber\\
& &\beta=6/g_0^2,\qquad 6.0\leq\beta\leq6.5,
\label{e_zpparam}
\ees
which is shown together with the data for $\zp$ in~\fig{f_zpparam}.
The formula describes the data with a statistical uncertainty of about
$0.5\%$.

\input zm_results_tab.tex

By combining the results for $M/\mbar$, $\za$ and $\zp$, we can now
compute $\zM$. Results at commonly used values of $\beta$ are listed
in~\tab{TAB_ZM}. The entries for $\zp$ and $\za$ in the table have
been taken from the parametrization \eq{e_zpparam} and Table~2 of
ref.~\cite{impr:pap4}, respectively. The first error quoted for $\zM$
comes from the uncertainty in the ratio $\za/\zp$, whereas the second
arises from the error in~$M/\mbar$ at the matching scale. It is useful
to separate the two sources of error, since $M/\mbar$ has been
computed in the continuum limit. Therefore, it does not make sense to
include the error quoted in~\eq{e_R1} in the continuum extrapolation
of lattice data for current quark masses. Instead the uncertainty of
1.3\% in $M/\mbar(\mu)$ at $\mu=(1.436\,r_0)^{-1}$ should be added in
quadrature to the quark mass {\it after\/} the extrapolation has been
performed.

As in the case of $\zp$, it is convenient to represent the results for
$\zM$ in terms of a polynomial fit function. For bare couplings in the
range $6.0\leq\beta\leq6.5$ the data for $\zM(g_0)$ are well described
by the expression
\bes
  \zM(g_0) = 1.752 +0.321\,(\beta-6) -0.220\,(\beta-6)^2\, .
\label{EQ_ZMparam}
\ees
This parametrization yields $\zM$ with an accuracy of about 1.1\%, if
the error in $M/\mbar$ is neglected.

As explained in detail in \App{a_zp} the results for $\zp$ are subject
to a systematic error of order~$a$, due to imperfect knowledge of the
improvement coefficient~$\ct$. Our estimates of this systematic error
show that it is negligible in most applications. In fact, the error in
the total renormalization $\zM$ is clearly dominated by the
uncertainty in $\za$, but not by residual $\Oa$ effects in $\zp$.

We now discuss briefly the application of our results for $\zM$.
Let us assume that we want to compute the sum of up and strange quark
masses, $M_{\rm u} + M_{\rm s}$. Using lattice data for the hadronic
scale $r_0$~\cite{pot:r0_SU3} and the bare current quark masses
computed in the $\Oa$ improved theory one can compute the
dimensionless quantity
\bes
\zM(g_0)\times(m_{\rm u}+m_{\rm s})(g_0)\times r_0(g_0).
\ees
Depending on the chosen values of the bare coupling, $\zM$ can be
taken either from~\tab{TAB_ZM} (using the first of the two quoted
errors) or the parametrization in~\eq{EQ_ZMparam} (with an uncertainty
of 1.1\%). An extrapolation in $(a/r_0)^2$ to the continuum limit
yields the value of $M_{\rm u} + M_{\rm s}$ in units of $r_0$. At this
point the error of 1.3\% in $M/\mbar$ at the matching scale should be
added in quadrature to the result. Estimates for the light quark
masses using our values of $\zM$ will be published
elsewhere~\cite{quark:alpha_prep}.

\subsection{$\zM$ for different lattice actions}

The results for $\zM$ presented here have been obtained in $\Oa$
improved lattice QCD using the action defined in~\cite{impr:pap3}. If
a different discretization of the QCD action were chosen, $\zM$ would
have to be recomputed. It is now important to realize that the
calculation of the universal part of $\zM$, i.e. the ratio $M/\mbar$
does not have to be repeated, since it is independent of the
regularization. This represents a considerable saving of CPU time,
since the most difficult and demanding part in the determination of
$\zM$ is the calculation of the scale dependence of $M/\mbar$ from low
to very high energies. Therefore, only the calculation of $\za$ (as
outlined in~\cite{impr:pap4}) and~$\zp$ has to be performed for a
different lattice action. 

%% file: zm_results_tab.tex
\begin{table}[bt]
\centering
\begin{tabular}{c c c c}
\hline \\[-1.0ex]
$\beta$ & $\zp$ & $\za$ & $\zM$ \\[1.0ex]
\hline \\[-1.0ex]
6.0 & 0.5233(26) & 0.7906(94) & 1.748(23)(23) \\
6.2 & 0.5178(26) & 0.8067(76) & 1.803(19)(23) \\
6.4 & 0.5157(26) & 0.8273(78) & 1.856(20)(24) \\[1.0ex]
\hline
\end{tabular}
\caption[TAB_ZM]{\footnotesize
Results for $\zp$, $\za$ and $\zM$ at typical values of
$\beta=6/g_0^2$
\label{TAB_ZM}
}
\end{table}

%% file: sect7.tex
\section{Conclusions \label{s_concl}}

In this paper we have demonstrated that the relation between the
fundamental parameters of QCD and hadronic observables can be computed
non-perturbatively with controlled errors. Our main results have been
obtained in the quenched approximation, but our methods carry over
literally to full QCD. That is, no additional conceptual difficulties
are expected, as long as one is interested in QCD with up, down and
strange quarks only.

By combining a recursive finite-size technique with lattice
simulations we have calculated the scale evolution of quark masses in
the \SF\ renormalization scheme, starting from energies of a few
hundred $\MeV$ up to scales well above 100\,\GeV, where contact with
perturbative scaling could be made. In this way we were able to
extract the ratio $M/\mbar$ at $\mu=(1.436\,r_0)^{-1}$. Similarly, by
computing the scale evolution of the running coupling we calculated
the $\Lambda$-parameter without compromising approximations. By means
of a careful removal of lattice artefacts in the computation of the
scale evolution we have obtained universal results for $M/\mbar$ and
the $\Lambda$-parameter.

Another main result of our work is the calculation of the
renormalization factor of the pseudoscalar density in $\Oa$ improved
lattice QCD for a range of bare couplings. We have computed this
factor in the SF scheme at the lowest energy matching point. The
results can be combined with the previously computed renormalization
factor of the axial current \cite{impr:pap4} and the universal factor
$M/\mbar$. Thereby we have determined the total renormalization
factor which relates the bare quark masses in the $\Oa$ improved
lattice theory to the renormalization group invariant quark masses.

Our results for $\Lambda_{\msbar}^{(0)}$ and the relation between the
bare current quark masses and $M$ are not subject to systematic errors
other than the use of the quenched approximation. In this sense our
method to compute renormalization factors of scale-dependent
quantities is rather unique. Another advantage is that renormalization
factors can be computed for vanishing quark mass. The method is
generally applicable to the problem of scale-dependent renormalization
and is now also being used to compute renormalization constants for
low moments of structure functions \cite{struct:roma2}.

Usually quark masses are quoted in the $\msbar$ scheme at
normalization mass $\mu=1\,\GeV$ or $\mu=2\,\GeV$. Once a
non-perturbative solution of the theory becomes possible, this
convention is not entirely satisfactory, because the $\msbar$ scheme
is only meaningful to any finite order of perturbation theory. On the
other hand, the renormalization group invariant quark masses are
non-perturbatively defined and scheme-independent. Hence, they are
more quotable than the $\msbar$ masses, and we would like to recommend
their use in future studies.

\vspace{1cm}
This work is part of the ALPHA collaboration research programme. We
are grateful to Roberto Petronzio for computer time on the
APE/Quadrics at the University of Rome ``Tor Vergata'', and to Giulia
de Divitiis and Tereza Mendes for their help.

\vfill
\eject

%% file: appendix1.tex
\section{Computation of step scaling functions on the lattice
\label{a_sigmas}} 

\subsection{Calculation of $\sigma(u)$ \label{a_sigma}}

The calculation starts from the lattice step scaling function
$\Sigma(u,a/L)$ defined as in the continuum but for finite
resolution $a/L$. Its precise definition was given in
ref.~\cite{alpha:SU3}. We followed closely this reference.  In
particular we took exactly the same discretization and set the
improvement coefficient $\ct$ to its value in one-loop perturbation
theory \cite{alpha:SU3},
\bes
  \ct^{\rm 1-loop} = 1 - 0.089 g_0^2\, . \label{e_ct1loop}
\ees
One then has to be aware that lattice spacing errors linear in the
lattice spacing are not suppressed completely, because perturbation
theory approximates $\ct$ with unknown precision.

Apart from numerical checks and a few simulations on the smallest
systems, the simulations were performed on APE-100 parallel computers
with 128 to 512 nodes. We employed the hybrid overrelaxation algorithm
described in Sect.~3.1 of \cite{impr:pap3}, setting $N_{\rm
OR}=L/(2a)$. For renormalized couplings of around $u=3.5$, we
used the modified sampling introduced in Appendix~A of
\cite{alpha:SU3} with parameter $\gamma=0.02\;-\;0.05$.

Numerical results for running couplings on pairs of lattices $L/a$,
$2L/a$ at the respective values of $\beta=6/g_0^2$, are listed in
\tab{TAB_sigma_results} for future reference. The data have been
analyzed as in ref.~\cite{alpha:SU3}, by propagating the error of
$\gbar^2(L)$ into $\Sigma(u,a/L)$, where $u$ is the central value of
$\gbar^2(L)$. We then extrapolated $\Sigma(u,a/L)$ to the continuum
limit, allowing for the expected dominant discretization error linear
in $a/L$. In practice we fit
\bes
 \Sigma(u,a/L) = \sigma(u) + \omega(u)\, a/L.
\label{e_Sigma}
\ees
These fits have good $\chi^2$. We list the fit parameters $\sigma(u)$,
the slopes $\omega(u)$ and the $\chi^2$ together with the number of
degrees of freedom in the fit, ${{n_{\rm df}}}$,
in~\tab{TAB_sig_extra}. One observes that at the level of our
precision lattice artefacts are not significant. Alternatively one
could assume that $\ct$ is well approximated by its one-loop
perturbative estimate in~\eq{e_ct1loop}, so that $a/L$ effects are
negligible. An extrapolation using a quadratic ansatz, $(a/L)^2$, for
the leading lattice artefacts would give entirely consistent values
for $\sigma(u)$ but smaller statistical errors.

In~\tab{TAB_sigma} we have included the results for $\sigma(u)$ from
\tab{TAB_sig_extra} and those in Table~4 of ref.~\cite{alpha:SU3}.

\input sigma_results_tab.tex

\input sig_extra_tab.tex

\subsection{Calculation of $\sigmap(u)$ \label{a_sigmap}}

We now describe the details of the numerical calculations which we
have performed to determine the step scaling function for $\zp$. This
involves the calculation of correlation functions constructed in the
framework of the \SF\ in a lattice simulation. Here we have used the
$\Oa$ improved Wilson action defined in \cite{impr:pap1}. We take over
the notation introduced in that reference.

\subsubsection{Definitions \label{a_sigmap_def}}

As explained in \sect{s_SF}, the renormalization constant $\zp$ is
defined in terms of correlation functions involving the pseudoscalar
density and the boundary quark fields. On the lattice these
correlation functions are represented by
\bes
  \fp(x_0) & = & -a^6\sum_{\vecy,\vecz}\frac{1}{3}\langle
         \psibar(x)\gamma_5\frac{1}{2}\tau^a\psi(x)
         \zetabar(\vecy)\gamma_5\frac{1}{2}\tau^a\zeta(\vecz)
         \rangle \label{EQ_fp} \,,
         \\
  f_1 & = & -{{a^{12}}\over{L^6}} \sum_{\vecu,\vecv,\vecy,\vecx}
         \frac{1}{3}\langle
         \zetabarprime(\vecu)\gamma_5\frac{1}{2}\tau^a\zetaprime(\vecv)
         \zetabar(\vecy)\gamma_5\frac{1}{2}\tau^a\zeta(\vecz)
         \rangle \label{EQ_f1}.
\ees
They have been formally defined in the continuum theory in
eqs. (\ref{e_fp},\ref{e_f1}), and we use the same symbols to denote
the lattice correlation functions.

The normalization constant $\zp$ of the pseudoscalar density is then
defined by
\be
  \zp(g_0,L/a) = c\,{\sqrt{3f_1}\over{\fp(L/2)}}
\label{EQ_zpnorm}
\ee
at vanishing quark mass. The constant $c$ is chosen so that $\zp=1$ at
tree level. For $\theta=1/2$, which is chosen in our simulations, we
list values for $c$ in~\tab{TAB_zptree}. They differ from unity by
small terms of order $(a/L)^2$.

\input zptree_tab.tex

We now need to specify the precise definition of the quark mass used
in the determination of $\zp$ and the step scaling function. The
reason is that the point of vanishing current quark mass is not
unambiguously defined in a regularization without exact chiral
symmetry. For instance, in the $\Oa$ improved theory there is an
uncertainty of order $a^2$. The details of the definition will
influence the size of $\Oasq$ lattice artefacts in the lattice step
scaling function, but not its continuum limit. Following
ref.~\cite{impr:pap3} we introduce the correlation function of the
(unimproved) bare axial current
\be
  \fa(x_0)  =  -a^6\sum_{\vecy,\vecz}\frac{1}{3}\langle
         \psibar(x)\gamma_0\gamma_5\frac{1}{2}\tau^a\psi(x)
         \zetabar(\vecy)\gamma_5\frac{1}{2}\tau^a\zeta(\vecz)
         \rangle, \label{EQ_fa}
\ee
and define an unrenormalized current quark mass~$m(g_0)$ through
\be
   m(g_0) = {\frac{1}{2}(\partial_0^*+\partial_0)\fa(x_0)
       +{\ca}a\partial_0^*\partial_0\fp(x_0) \over 2\fp(x_0)}
   \;.
\label{EQ_mPCAC}
\ee
Here, $\partial_0$ and $\partial_0^*$ are the forward and backward
lattice derivatives, respectively, and $\ca$ denotes the coefficient
multiplying the $\Oa$ improvement term in the improved axial current.
It is understood that~\eq{EQ_mPCAC} is evaluated for $\theta=0$,
whereas in all other cases the correlation functions have been
computed for $\theta=1/2$.

The lattice step scaling function $\sigplatt$ is then defined through
\be
  \sigplatt(u,a/L) = 
  {{\zp(g_0,2L/a)}\over{\zp(g_0,L/a)}}
  \quad\hbox{at}\;\;m(g_0)=0,\;\;\gbar^2(L)=u.
\label{EQ_sigplatt}
\ee
Here, $m(g_0)$ is always evaluated for lattice size $L/a$ with
$x_0=L/2$. The condition $m(g_0)=0$ defines the critical value of the
hopping parameter, $\hopc$, and the condition $\gbar^2(L)=u$ specifies
which value of the bare coupling $g_0$ is to be used for a given value
of $L/a$.

With these definitions the lattice step scaling function $\sigplatt$ is
a function of the renormalized coupling $u$, up to cutoff effects. Its
continuum limit is reached as $a/L \to 0$ for fixed $u$.

\subsubsection{Details of the simulation}

The correlation functions in eqs.~(\ref{EQ_fp}), (\ref{EQ_f1})
and~(\ref{EQ_fa}) have been evaluated in a lattice simulation using
the $\Oa$ improved Wilson action and the axial current defined in
\cite{impr:pap3}. In particular, the improvement coefficients $\csw$
and $\ca$ given by eqs.~(5.16) and~(6.5) of that reference have been
used.

As in the case of $\Sigma$, boundary $\Oa$ improvement terms have to
be considered. In analogy to $\ct$, one also has to cancel the
contributions of boundary terms arising from local composite operators
involving quark fields. This is achieved by including a suitably
chosen term in the action, whose coefficient is denoted by $\cttil$
\cite{impr:pap1}. Also $\cttil$ has only been determined in
perturbation theory. We have used its perturbative expansion to one
loop \cite{impr:pap2}
\be
   \cttil^{\rm 1-loop} = 1-0.018g_0^2\, .
\label{EQ_cttilde}
\ee
The influence of the imperfectly known improvement coefficients $\ct$
and $\cttil$ on the continuum extrapolation has to be investigated. In
particular, it is not clear a priori whether an extrapolation of
$\sigplatt$ in $a^2/L^2$ is justified.

The constants $\zp(g_0,L/a)$ and $\zp(g_0,2L/a)$ in~\eq{EQ_sigplatt}
have been evaluated for $\theta=1/2$ at the point where the current
quark mass defined in~\eq{EQ_mPCAC} vanishes. The point in $\hop$
where the condition $m(g_0)=0$ is satisfied is called the ``critical''
value of the hopping parameter, $\hopc$. In practice, $\hopc$ has been
obtained through an interpolation of the data for $m(g_0)$ around the
point where it vanishes.

We have evaluated $\sigplatt(u,a/L)$ at each of the 13 chosen values
of $u$ for $L/a=6,8,12$ and~16. The simulation algorithm is as
specified before and the correlation functions were computed as
detailed in Sect.~2 and Subsect.~3.2 of
ref.~\cite{impr:pap3}. ``Measurements'' of the correlation functions
$f_1$, $\fa$ and $\fp$ were separated by 5 full iterations of the
algorithm on the smallest lattice, rising to 30 iterations on the
largest. We have checked explicitly for the statistical independence
of our sample by dividing the full ensemble into bins, each containing
a number of individual ``measurements''. The statistical errors were
then monitored as the number of measurements per bin was increased. We
did not observe any significant change of the errors in $\zp$ for
increasing bin size, which we take as evidence for the statistical
independence of all our samples.

Since boundary conditions $C=C^\prime=0$ apply here, the correlation
functions $\fa$ and $\fp$ could be averaged with their counterparts
$\fa^\prime$ and $\fp^\prime$, which are defined by a time-reflection
applied to $\fa$ and $\fp$ (see eqs.~(2.5) and~(2.6) of
ref.~\cite{impr:pap3}).

The number of measurements was chosen such that the statistical error
in $\zp(g_0,L/a)$ was typically a factor 2--3 smaller than that of
$\zp(g_0,2L/a)$. Thereby it was ensured that the statistical error of
$\sigplatt$ was dominated by the uncertainty in
$\zp(g_0,2L/a)$. Typically, we have evaluated $\zp(g_0,L/a)$ on
$200-640$ configurations, whereas $60-120$ measurements were
accumulated for $\zp(g_0,2L/a)$.

Statistical errors were computed using the jackknife method. The error
in $\sigplatt$ due to the uncertainty in the coupling was estimated
using the one-loop expansion of $\sigplatt$ in~$u$. It turned out to
be about 10 times smaller than the statistical error in $\sigplatt$ in
the whole range of~$u$ considered and has therefore been neglected
in the error estimate. The error in $\sigplatt$ due to the uncertainty
in $\hopc$ has been estimated using the slopes for $\zp(g_0,L/a)$ and
$\zp(g_0,2L/a)$ computed for several values of the hopping
parameter. It was found that $\sigplatt$ depends rather weakly on the
bare quark mass, so that the uncertainty in $\sigplatt$ from the error
in $\hopc$ was negligible.

Results for $\zp(g_0,L/a)$, $\zp(g_0,2L/a)$, $\hopc$ and the step
scaling function $\sigplatt$ are shown in \tab{TAB_sigp_results}. The
data in the last four rows were computed using the two-loop formula
for $\ct$ (see \eq{e_ct2loop} below).

\subsubsection{Continuum extrapolation of $\sigplatt$}

The next task is the extrapolation of $\sigplatt$ to the continuum
limit for fixed coupling~$u$. Our results for $\sigplatt$ show at most
a weak dependence on the lattice spacing. Before performing an
extrapolation we first investigate the question whether it is
justified to assume that the leading cutoff effects are proportional
to $a^2/L^2$, given that the improvement coefficients $\ct$ and
$\cttil$ are only known in perturbation theory.

We start by investigating the influence of $\cttil$ by comparing the
step scaling function computed using the one-loop expression for
$\cttil$ to the case where we artificially replace the one-loop
coefficient by 10 times its value in \eq{EQ_cttilde}, viz.
\be
   \cttil^{\,\prime} = 1-0.180\,g_0^2\,.
\ee
For the range of bare couplings used in our simulations, this
represents a change of up to 20\%, which is surely a conservative
estimate of the remainder of the perturbative series for $\cttil$. We
found that the change in $\sigplatt$ induced by replacing $\cttil^{\rm
1-loop}$ by $\cttil^{\,\prime}$ amounts to around 0.8\% for $L/a=6$
and $u=3.48$, i.e. at the point where the effect is expected to be
most pronounced. Furthermore, the difference in $\sigplatt$ decreases
considerably for $L/a=8$ and/or smaller couplings. We conclude that
the step scaling function is only weakly dependent on the value of
$\cttil$, and that the effect of using the one-loop estimate for
$\cttil$ is negligible at the current level of precision.

To assess the influence of $\ct$ on the cutoff dependence of
$\sigplatt$, one has to take into account that the bare coupling must
be adjusted when $\ct$ is changed in order to keep the renormalized
coupling fixed. We have performed this analysis for fixed
$u=3.48$ where the uncertainty in $\ct$ is largest. We changed
$\ct$ by adding the two-loop term, which was only known towards the
end of our numerical computations~\cite{pert:2loop},
\be
  \ct^{\rm 2-loop} = 1 - 0.089 g_0^2-0.030 g_0^4\, . \label{e_ct2loop}
\ee
The corresponding change in $\Sigmap$ is $\approx1.5\%$ for $L/a=6$,
decreasing considerably as the resolution gets finer
(cf. \tab{TAB_sigp_results}). We will show in the following that this
$\Oa$-uncertainty, although statistically significant at non-zero $a/L$,
does not affect our values for $\sigmap(u)$ obtained by continuum
extrapolations.  

\input sigp_results_tab.tex

The dependence of $\sigplatt$ on the lattice spacing is modelled using
the following fit functions
\bes
  {\rm Fit~A:}&\quad&\sigplatt(u,a/L) = \sigpcont(u)
  +\rho(u)\,a/L \\
  {\rm Fit~B:}&\quad&\sigplatt(u,a/L) = \sigpcont(u)
  +\rho^\prime(u)\,a^2/L^2.
\ees
In other words, we compare the approach to the continuum limit
assuming that the leading cutoff effects are either linear (Fit~A) or
quadratic (Fit~B) in $a/L$. Fit~A is motivated by the fact that not
all $\Oa$ improvement terms are known non-perturbatively.  As a
safeguard against higher order cutoff effects, we exclude the data for
$\sigplatt$ obtained on our coarsest lattices, i.e. for $L/a=6$ from
the fits.

\begin{figure}[tb]
\hspace{0cm}
\vspace{-0.0cm}
 
\centerline{
\psfig{file=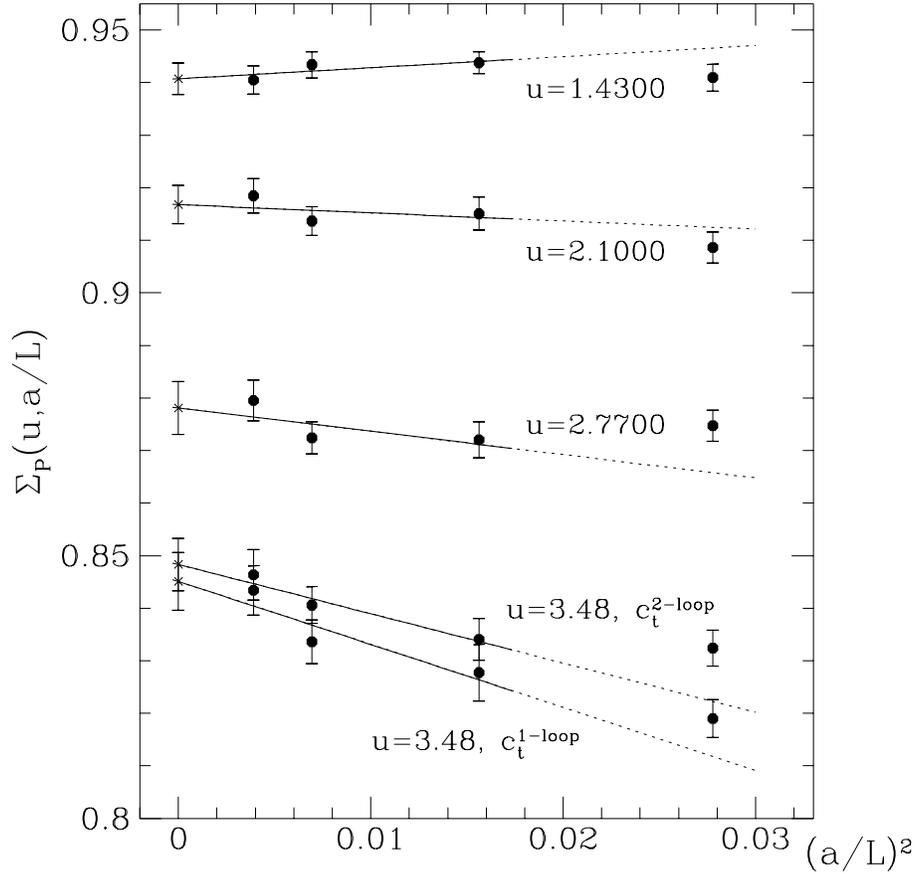,width=14cm}
}
\vspace{1.5cm}
\caption{\footnotesize
 Examples of continuum extrapolations of $\sigplatt$ using Fit~B. The
 dotted lines are the continuation of the fit functions to the data
 points for $L/a=6$, which have been excluded from the fit.
\label{FIG_sigp_extra}
}
\end{figure}

The extrapolated values for $\sigmap(u)$ show no significant
dependence on the fit ansatz (Fit~A or~B) for all renormalized
couplings~$u$. The biggest effect is seen at the largest coupling,
i.e. $u=3.48$, where the results for $\sigpcont(3.48)$ obtained from
either Fit~A or~B differ by one standard deviation.  Such an effect
may be statistical or systematic. To test for the latter possibility,
we consider the data for $\Sigmap(3.48,a/L)$ obtained using
$\ct=\ct^{\rm 2-loop}$. \Fig{FIG_sigp_extra} demonstrates that cutoff
effects of $\Oa$ are reduced compared to $\ct=\ct^{\rm
1-loop}$. Furthermore, Fit~B applied to the data set for $\ct=\ct^{\rm
2-loop}$ produces an extrapolated value which is entirely consistent
with the one obtained for $\ct=\ct^{\rm 1-loop}$. We conclude that the
small uncertainties which are present in the improvement coefficients
$\ct,\cttil$ are numerically unimportant, and that extrapolations
using $(a/L)^2$ terms as the dominant scaling violation are justified.
We emphasize that such a statement can only be made for a given level
of statistical accuracy. Furthermore we use only data for $L/a=8$, 12
and~16, with the last point already fairly close to the continuum
limit.

For our best estimates we have taken the results from Fit~B, obtained
for the standard one-loop result for $\ct$, and excluding the data for
$L/a=6$. Typical examples of extrapolations at selected values of~$u$
are shown in \fig{FIG_sigp_extra}. The fit parameters for all
extrapolations using Fit~B are shown in \tab{TAB_sigp_extra}, where
the point computed using the two-loop expression for the improvement
coefficient $\ct$ is marked by an asterisk.

\input sigp_extra_tab.tex

%% file: sigma_results_tab.tex
\begin{table}
\centering
\begin{tabular}{rrlll}
\hline \\[-1.0ex]
   $\beta~~~$ &  $L/a$ & $~\gbar^2(L)$ & $~\gbar^2(2L)$ 
& $\Sigma(u,a/L)$ \\[1.0ex]
\hline \\[-1.0ex]
 10.6064 & 5&  0.8873(5) &  0.9736(28) &
 0.9736(29) \\
 10.7503 & 6&  0.8873(5) &  0.9712(29) &
 0.9712(30) \\
 10.8790 & 7&  0.8873(5) &  0.9767(31) &
 0.9767(32) \\
 11.0000 & 8&  0.8873(10)&  0.9759(28) &
 0.9759(31)
  \\[1.0ex]
  9.9024 & 5&  0.9944(6) &  1.1070(32) &
 1.1070(33) \\
 10.0500 & 6&  0.9944(7) &  1.1044(38) &
 1.1044(39) \\
 10.1835 & 7&  0.9944(7) &  1.1024(38) &
 1.1024(39) \\
 10.3000 & 8&  0.9944(13)&  1.1127(41) &
 1.1127(44)
  \\[1.0ex]
  9.3518 & 5&  1.0989(7) &  1.2318(41) &
 1.2318(42) \\
  9.5030 & 6&  1.0989(8) &  1.2350(46) &
 1.2350(47) \\
  9.6272 & 7&  1.0989(8) &  1.2434(35) &
 1.2434(36) \\
  9.7500 & 8&  1.0989(13)&  1.2364(34) &
 1.2364(38)
  \\[1.0ex]
  8.4582 & 5&  1.3293(10) & 1.5541(62) &
 1.5541(63) \\
  8.6129 & 6&  1.3293(12) & 1.5409(52) &
 1.5409(54) \\
  8.7431 & 7&  1.3293(11) & 1.5442(76) &
 1.5442(77) \\
  8.8500 & 8&  1.3293(21) & 1.5461(70) &
 1.5461(76)
  \\[1.0ex]
  7.8583 & 5&  1.5553(13) & 1.8776(93) &
 1.8776(95) \\
  7.9993 & 6&  1.5553(15) & 1.8811(38) &
 1.8813(43) \\
  8.1380 & 7&  1.5553(19) & 1.884(11)&
 1.884(12) \\
  8.2500 & 8&  1.5553(24) & 1.864(10) &
 1.864(11) \\
  8.5873 &12&  1.5553(50) & 1.879(17)&
 1.879(19)
  \\[1.0ex]
  7.2611 & 5&  1.8811(19) & 2.388(15)&
 2.388(15) \\
  7.4082 & 6&  1.8811(22) & 2.397(17)&
 2.397(18) \\
  7.5438 & 7&  1.8811(26) & 2.374(18)&
 2.374(18) \\
  7.6547 & 8&  1.8811(28) & 2.393(18)&
 2.393(18) \\
  7.9993 &12&  1.8811(38) & 2.411(20)&
 2.411(21)
  \\[1.0ex]
  6.6255 & 5&  2.4484(32) & 3.504(16)&
 3.504(17) \\
  6.7807 & 6&  2.4484(37) & 3.478(22)&
 3.478(23) \\
  6.9079 & 7&  2.4484(56) & 3.501(17)&
 3.501(21) \\
  7.0197 & 8&  2.4484(45) & 3.484(21)&
 3.484(23) \\
  7.1190 & 9&  2.4484(58) & 3.496(42)&
 3.496(44) \\
  7.3551 &12&  2.4484(80) & 3.475(26)&
 3.475(31)
  \\[1.0ex]
\hline
\end{tabular}
\caption[TAB_sigma_results]{\footnotesize
Pairs of renormalized couplings
and the step scaling function $\Sigma$
\label{TAB_sigma_results}
}
\end{table}
\clearpage

%% file: sig_extra_tab.tex
\begin{table}[tb]
\centering
\begin{tabular}{l l r@{.}l c}
\hline \\[-1.0ex]
\multicolumn{1}{c}{$u$}  &  $\sigma(u)$
 & \multicolumn{2}{c}{$\omega(u)$} &
$\chi^2/{{n_{\rm df}}}$ \\[1.0ex]
\hline \\[-1.0ex]
0.8873   &       0.981(9)  &    $-0$&04(5)   &    1.3/2 \\
0.9944   &       1.110(11)  &   $-0$&02(7)   &   3.3/2 \\
1.0989   &       1.252(11)  &   $-0$&10(7)   &    3.0/2 \\
1.3293   &       1.528(20)  &      0&11(12)  &   1.7/2 \\
1.5553   &       1.865(23)  &      0&09(14)  &   2.4/4 \\
1.8811   &       2.412(32)  &   $-0$&14(21)  &   1.5/3 \\
2.4484   &       3.464(40)  &      0&19(25)  &   0.9/4
  \\[1.0ex]
\hline
\end{tabular}
\caption[TAB_sig_extra]{\footnotesize 
Continuum extrapolation of the step scaling function $\Sigma$ according
to \eq{e_Sigma} 
\label{TAB_sig_extra}
}
\end{table}

%% file: zptree_tab.tex
\begin{table}[b]
\centering
\begin{tabular}{rccrc}
\hline \\[-1.0ex]
$L/a$ & $c$ & & $L/a$ & $c$\\[1.0ex]
\hline \\[-1.0ex]
  6 & 0.98969662  & & 16 & 0.99853742  \\
  8 & 0.99417664  & & 24 & 0.99934941  \\
 12 & 0.99740297  & & 32 & 0.99963393  \\[1.0ex]
\hline
\end{tabular}
\caption[TAB_zptree]{\footnotesize
Values for $c$, defined implicitly by $\zp(0,L/a)=1$ 
\label{TAB_zptree}
}
\end{table}

%% file: sigp_results_tab.tex
\begin{table}
\centering
\begin{tabular}{rlrlllc}
\hline \\[-1.0ex]
   $\beta~~~$ & $~~~\hopc$ & $L/a$ & $~~~\gbar^2(L)$ & $\zp(g_0,L/a)$
 & $\zp(g_0,2L/a)$ & $\sigplatt(u,a/L)$  \\[1.0ex]
\hline \\[-1.0ex]
10.7503 & 0.130591(4) & 6 & 0.8873(5)  & 0.8476(5) & 0.8187(13) &
   0.9659(17) \\
11.0000 & 0.130439(3) & 8 & 0.8873(10) & 0.8407(5) & 0.8118(14) &
   0.9656(18) \\
11.3384 & 0.130251(2) &12 & 0.8873(30) & 0.8314(6) & 0.8035(10) &
   0.9664(14) \\
11.5736 & 0.130125(2) &16 & 0.8873(25) & 0.8252(8) & 0.7993(16) &
   0.9687(21)
  \\[1.0ex]
10.0500 & 0.131073(5) & 6& 0.9944(7)   & 0.8334(5) & 0.8010(13) &
   0.9612(17) \\
10.3000 & 0.130889(3) & 8& 0.9944(13)  & 0.8259(6) & 0.7950(16) &
   0.9625(21) \\
10.6086 & 0.130692(2) &12& 0.9944(30)  & 0.8143(7) & 0.7868(13) &
   0.9662(18) \\
10.8910 & 0.130515(2) &16& 0.9944(28)  & 0.8094(8) & 0.7813(13) &
   0.9653(19)
  \\[1.0ex]
 9.5030 & 0.131514(5) & 6 & 1.0989(8)  & 0.8190(6) & 0.7822(13) &
   0.9551(18) \\
 9.7500 & 0.131312(3) & 8 & 1.0989(13) & 0.8114(6) & 0.7778(16) &
   0.9586(21) \\
10.0577 & 0.131079(3) &12 & 1.0989(40) & 0.8007(7) & 0.7688(12) &
   0.9602(17) \\
10.3419 & 0.130876(2) &16 & 1.0989(44) & 0.7965(9) & 0.7661(16) &
   0.9617(22)
  \\[1.0ex]
 8.8997 & 0.132072(9) & 6& 1.2430(13)  & 0.8015(6) & 0.7638(16) &
   0.9529(21) \\
 9.1544 & 0.131838(4) & 8& 1.2430(14)  & 0.7935(7) & 0.7557(16) &
   0.9523(23) \\
 9.5202 & 0.131503(3) &12& 1.2430(35)  & 0.7853(7) & 0.7498(14) &
   0.9548(20) \\
 9.7350 & 0.131335(3) &16& 1.2430(34)  & 0.7769(9) & 0.7436(20) &
   0.9572(28)
  \\[1.0ex]
 8.6129 & 0.132380(6) & 6& 1.3293(12)  & 0.7917(6) & 0.7489(17) &
   0.9460(22) \\
 8.8500 & 0.132140(5) & 8& 1.3293(21)  & 0.7811(7) & 0.7423(17) &
   0.9503(23) \\
 9.1859 & 0.131814(3) &12& 1.3293(60)  & 0.7736(8) & 0.7344(15) &
   0.9494(22) \\
 9.4381 & 0.131589(2) &16& 1.3293(40)  & 0.7689(10)& 0.7282(17) &
   0.9470(25)
  \\[1.0ex]
 8.3124 & 0.132734(10)& 6& 1.4300(20)  & 0.7808(8) & 0.7347(19) &
   0.9410(26) \\
 8.5598 & 0.132453(5) & 8& 1.4300(21)  & 0.7713(7) & 0.7279(15) &
   0.9438(21) \\
 8.9003 & 0.132095(3) &12& 1.4300(50)  & 0.7627(8) & 0.7195(18) &
   0.9434(25) \\
 9.1415 & 0.131855(3) &16& 1.4300(58)  & 0.7562(11)& 0.7112(18) &
   0.9405(27)
  \\[1.0ex]
 7.9993 & 0.133118(7) & 6& 1.5553(15)  & 0.7663(7) & 0.7181(20) &
   0.9371(28) \\
 8.2500 & 0.132821(5) & 8& 1.5553(24)  & 0.7591(7) & 0.7128(20) &
   0.9390(27) \\
 8.5985 & 0.132427(3) &12& 1.5533(70)  & 0.7483(9) & 0.7046(14) &
   0.9415(22) \\
 8.8323 & 0.132169(3) &16& 1.5533(70)  & 0.7424(11)& 0.6939(22) &
   0.9346(32)
  \\[1.0ex]
\hline
\end{tabular}
\caption[TAB_sigp_results]{\footnotesize
Results for the step scaling function $\sigplatt$
\label{TAB_sigp_results}
}
\end{table}

\addtocounter{table}{-1}

\begin{table}
\centering
\begin{tabular}{rlrlllc}
\hline \\[-1.0ex]
   $\beta~~~$ & $~~~\hopc$ & $L/a$ & $~~~\gbar^2(L)$ & $\zp(g_0,L/a)$
 & $\zp(g_0,2L/a)$ & $\sigplatt(u,a/L)$  \\[1.0ex]
\hline \\[-1.0ex]
 7.7170 & 0.133517(8) & 6& 1.6950(26)  &  0.7520(7) & 0.7028(18) &
   0.9346(26) \\
 7.9741 & 0.133179(5) & 8& 1.6950(28)  &  0.7454(8) & 0.6972(19) &
   0.9354(28) \\
 8.3218 & 0.132756(4) &12& 1.6950(79)  &  0.7349(9) & 0.6851(14) &
   0.9322(22) \\
 8.5479 & 0.132485(3) &16& 1.6950(90)  &  0.7281(11)& 0.6779(20) &
   0.9312(31)
  \\[1.0ex]
 7.4082 & 0.133961(8) & 6& 1.8811(22)  &  0.7344(8) & 0.6787(19) &
   0.9241(28) \\
 7.6547 & 0.133632(6) & 8& 1.8811(28)  &  0.7268(8) & 0.6693(22) &
   0.9209(32) \\
 7.9993 & 0.133159(4) &12& 1.8811(38)  &  0.7187(8) & 0.6641(17) &
   0.9241(26) \\
 8.2415 & 0.132847(3) &16& 1.8811(99)  &  0.7104(11)& 0.6588(20) &
   0.9274(32)
  \\[1.0ex]
 7.1214 & 0.134423(9) & 6& 2.1000(39)  &  0.7168(8) & 0.6513(20) &
   0.9086(29) \\
 7.3632 & 0.134088(6) & 8& 2.1000(45)  &  0.7073(8) & 0.6472(21) &
   0.9151(31) \\
 7.6985 & 0.133599(4) &12& 2.1000(80)  &  0.6969(10)& 0.6368(16) &
   0.9137(27) \\
 7.9560 & 0.133229(3) &16& 2.100(11)   &  0.6930(12)& 0.6365(20) &
   0.9185(33)
  \\[1.0ex]
 6.7807 & 0.134994(11)& 6& 2.4484(37)  &  0.6874(10)& 0.6117(21) &
   0.8899(32) \\
 7.0197 & 0.134639(7) & 8& 2.4484(45)  &  0.6809(10)& 0.6079(21) &
   0.8928(33) \\
 7.3551 & 0.134141(5) &12& 2.4484(80)  &  0.6685(10)& 0.6004(18) &
   0.8981(30) \\
 7.6101 & 0.133729(4) &16& 2.448(17)   &  0.6688(12)& 0.5956(20) &
   0.8905(34)
  \\[1.0ex]
 6.5512 & 0.135327(12)& 6& 2.770(7)    &  0.6629(11)& 0.5799(18) &
   0.8747(30) \\
 6.7860 & 0.135056(8) & 8& 2.770(7)    &  0.6572(10)& 0.5731(21) &
   0.8720(35) \\
 7.1190 & 0.134513(5) &12& 2.770(11)   &  0.6496(10)& 0.5667(18) &
   0.8724(31) \\
 7.3686 & 0.134114(3) &16& 2.770(14)   &  0.6448(11)& 0.5671(23) &
   0.8795(39)
  \\[1.0ex]
 6.2204 & 0.135470(15)& 6& 3.480(8)    &  0.6181(12)& 0.5063(20) &
   0.8190(36) \\
 6.4527 & 0.135543(9) & 8& 3.480(14)   &  0.6114(10)& 0.5061(32) &
   0.8277(54) \\
 6.7750 & 0.135121(5) &12& 3.480(39)   &  0.6089(11)& 0.5076(24) &
   0.8336(42) \\
 7.0203 & 0.134707(4) &16& 3.480(21)   &  0.6063(12)& 0.5114(26) &
   0.8434(47)
  \\[1.0ex]
 6.257~ & 0.135499(14)& 6& 3.480(8)    &  0.6221(11)& 0.5179(19) &
   0.8324(34) \\
 6.476~ & 0.135524(9) & 8& 3.480(8)    &  0.6166(10)& 0.5143(23) &
   0.8341(40) \\
 6.799~ & 0.135090(6) &12& 3.480(9)    &  0.6106(11)& 0.5133(19) &
   0.8406(35) \\
 7.026~ & 0.134701(4) &16& 3.480(13)   &  0.6070(13)& 0.5137(27) &
   0.8464(48)
  \\[1.0ex]
\hline
\end{tabular}
\caption{\footnotesize (continued)}
\end{table}
\clearpage

%% file: sigp_extra_tab.tex
\begin{table}[tb]
\centering
\begin{tabular}{l l r@{.}l c}
\hline \\[-1.0ex]
\multicolumn{1}{c}{$u$}  &  $\sigpcont(u)$
 & \multicolumn{2}{c}{$\rho^\prime(u)$} &
$\chi^2/{{n_{\rm df}}}$ \\[1.0ex]
\hline \\[-1.0ex]
0.8873   &       0.9683(21)  &    $-0$&19(21)   &    0.51 \\
0.9944   &       0.9672(23)  &    $-0$&28(23)   &    0.47 \\
1.0989   &       0.9622(24)  &    $-0$&24(24)   &    0.09 \\
1.2430   &       0.9579(29)  &    $-0$&36(28)   &    0.13 \\
1.3293   &       0.9470(28)  &      0&23(28)    &    0.27 \\
1.4300   &       0.9407(30)  &      0&21(27)    &    0.40 \\
1.5553   &       0.9382(33)  &      0&11(33)    &    3.01 \\
1.6950   &       0.9297(32)  &      0&36(33)    &    0.00 \\
1.8811   &       0.9284(36)  &    $-0$&50(37)   &    0.21 \\
2.1000   &       0.9168(37)  &    $-0$&16(37)   &    1.10 \\
2.4484   &       0.8942(38)  &    $-0$&01(39)   &    3.04 \\
2.7700   &       0.8781(42)  &    $-0$&44(41)   &    1.44 \\
3.48     &       0.8451(55)  &    $-1$&20(59)   &    1.05 \\
3.48$^*$ &       0.8483(50)  &    $-0$&94(48)   &    0.26
  \\[1.0ex]
\hline
\end{tabular}
\caption[TAB_sigp_extra]{\footnotesize 
Continuum extrapolations of $\sigplatt$ using Fit B
\label{TAB_sigp_extra}
}
\end{table}

%% file: appendix2.tex
\section{Error propagation in the scale evolution \label{APX_sigma_fit}}

In this appendix we provide further details
about the numerical solution of the recursion relations 
(\ref{u_recursion}) and (\ref{v_recursion}), the principal aim being
to explain how precisely the errors on the final results 
have been determined.

As discussed in \Sect{s_runn}, the calculation starts by
fitting the data for the step scaling functions listed in \Tab{TAB_sigma}.
In the case of the function $\sigma(u)$ our fit ansatz is  
\be
  \sigma(u)=u+\sigma_0u^2+\sigma_1u^3+\ldots+\sigma_nu^{n+2}, 
\ee
where $\sigma_2,\ldots,\sigma_n$ are the fit parameters while
\be
  \sigma_0=2\ln(2)b_0
  \quad\hbox{and}\quad
  \sigma_1=\sigma_0^2+2\ln(2)b_1
\ee
are fixed to the values that one obtains in perturbation theory.
The fit function for the other step scaling function is taken to be
\be
  \sigmap(u)=1+\sigma_{{\rm P},0}u+
  \ldots+\sigma_{{\rm P},n}u^{n+1}, 
  \qquad
  \sigma_{{\rm P},0}=-\ln(2)d_0.
\ee
Least-squares fits with $1,2$ and $3$ fit parameters have then been
performed and were found to represent the data well (the curves shown
in Figs.~\ref{FIG_sigma_fit} and \ref{FIG_sigmap_fit} are for $2$ fit
parameters).

The solution of the recursion relations
(\ref{u_recursion}) and (\ref{v_recursion}) is unique
once a definite expression for the step scaling functions
is chosen.
In particular, the calculated
values of $u_k$ and $v_k$ are functions of the fit parameters.
However, one should take into account that 
the data for the step scaling functions 
have errors and thus determine the fit parameters only
within certain error margins. The errors on the step scaling functions 
are uncorrelated (they come from different simulation runs)
and the errors on the fit parameters and 
the $u_k$'s and $v_k$'s are thus obtained straightforwardly 
by applying the usual rules for error propagation.

The question now arises to which extent the calculated
values and error bounds are influenced
by the choice of fit functions.
We have made two independent checks to convince ourselves 
that this source of systematic error is under control.
First we noted that compatible results with slightly larger errors 
are obtained if the number of fit parameters is increased.
This is the expected behaviour, and since the fit quality is already good 
with one parameter, we decided to quote results for two fit parameters
so as to be on the safe side.

The other check that we have made is to apply the fit procedure
and error determination to simulated data sets that have been 
generated artificially, assuming $\sigma(u)$ and $\sigmap(u)$
to be equal to some analytically given functions with 
roughly the right shape.
Comparing with the ``exact'' solution of the recursion relations,
we then found that the systematic bias arising from 
the choice of fit polynomials is not significant at the current level of 
precision.

%% file: appendix3.tex
\section{Computation of $\zp$ at the matching scale \label{a_zp}}

Here we describe the details of the calculation of
$\zp(g_0,L/a)_{L=1.436\,\rnod}$ required to connect the masses in the
SF-scheme to the bare current quark masses. Our task is to compute
$\zp$ for a range of bare couplings, which are commonly used in
simulations in large physical volumes. In addition, we have to
estimate the effect of using perturbation theory for the improvement
coefficients $\ct$ and $\cttil$, instead of non-perturbative values,
which are currently not available. Since $\ct$ is known to two loops,
we have computed $\zp$ using both the one- and two-loop expressions,
in order to assess the influence on the final estimates.

Our procedure is as follows. First we choose pairs of $(\beta,L/a)$
such that $L/a=1.436\,r_0/a$, using lattice data for $r_0/a$ from
\cite{pot:r0_SU3}. For a chosen value of $L/a$ this condition determines
the corresponding $\beta$-value, which is obtained by inserting
$a/r_0=1.436(a/L)$ into the parametrization of $\ln(a/r_0)$ quoted in
eq.~(2.18) of ref.~\cite{pot:r0_SU3}, viz.
\be
 \ln(a/r_0) = -1.6805  -1.7139\,(\beta-6)
  +0.8155\,(\beta-6)^2 -0.6667\,(\beta-6)^3.
\label{e_poly_fit}
\ee
Solving numerically for $\beta$ yields the desired combination of
$(\beta,\,L/a)$. $\zp$ is then computed for $\theta=0.5$ using the
normalization condition \eq{EQ_zpnorm} at the critical value of the
hopping parameter determined for lattice size $L/a$.

This procedure yields $\beta$ with finite accuracy, and the error
which propagates into $\zp$ can be estimated using one-loop
perturbation theory. This error can be neglected, since it is more
than 10 times smaller than the statistical error in $\zp$. Results for
$\zp(g_0,L/a)_{L=1.436\,r_0}$ computed using both the one- and
two-loop expressions for $\ct$ are listed in \tab{TAB_zp}.

\input zp_tab.tex

In principle one could compute $\zp$ for smaller values of $\beta$ and
$L/a$. However, for $L/a=6$ the condition $L=1.436\,r_0$ implies
$\beta<6.0$, where the parametrizations of $\csw$ and $\ca$ from
ref.~\cite{impr:pap3} cannot be used. In order to have an additional
value at $\beta=6.0$, we have applied a special procedure. We have
computed $\zp$ for $\beta=6.0$ fixed, but using different lattice
sizes for which $L/r_0$ straddles the reference value
$L/r_0=1.436$. An interpolation in $L/r_0$ at fixed $\beta$ then
yields the desired result for $\zp(g_0=1,L/a)_{L=1.436\,r_0}$.

Using the parametrization of $r_0/a$, eq.~(2.18) in \cite{pot:r0_SU3} we
find
\be
   r_0/a = 5.368(22),\quad\beta=6.0,
\ee
and the results for $\zp$ for three different values of $L/r_0$ are
listed in \tab{TAB_zp60}.

\input zp60_tab.tex

A quadratic interpolation in $L/r_0$ of the data in the table yields
\be
   \zp(g_0=1,L/a)_{L=1.436\,r_0} = \left\{\begin{array}{lr}
	0.5279(24), & \ct^{\rm 1-loop} \\
	0.5253(26), & \ct^{\rm 2-loop}
	\end{array}\right. .
   \label{e_zp_beta6}	
\ee
The results for $\zp(g_0,L/a)_{L=1.436\,r_0}$ for $6.0\leq
\beta\leq6.5$, using either $\ct^{\rm 1-loop}$ or $\ct^{\rm 2-loop}$
can be parametrized by a polynomial fit in $(\beta-6)$. For $\ct^{\rm
2-loop}$ the result is shown in \eq{e_zpparam}.

We end this appendix with a brief comment on the influence of
$\cttil$ on the estimates of $\zp$. Similarly to the case of
$\sigplatt$, this has been investigated by repeating the computation
of $\zp$ at $\beta=6.0$ using $\cttil$ with its one-loop coefficient
multiplied by a factor 10. It turned out that the resulting
interpolated value of $\zp(1,L/a)_{L=1.436\,r_0}$ did not change
appreciably within statistical errors. We conclude that the influence
of $\cttil$ can be neglected at the present level of precision.

%% file: zp_tab.tex
\begin{table}[tb]
\centering
\begin{tabular}{r c c c c}
\hline \\[-1.0ex]
      &         & $\ct^{\rm 1-loop}$ & $\ct^{\rm 2-loop}$ &  \\
$L/a$ & $\beta=6/g_0^2$ & $\zp$ & $\zp$ \\[1.0ex]
\hline \\[-1.0ex]
 8 & 6.0219 & 0.5250(17)  &    0.5218(17)   \\
10 & 6.1628 & 0.5211(17)  &    0.5177(19)  \\
12 & 6.2885 & 0.5207(21)  &    0.5179(23)  \\
16 & 6.4956 & 0.5160(17)  &    0.5157(19)  \\[1.0ex]
\hline
\end{tabular}
\caption[TAB_zp]{\footnotesize
Results for $\zp(g_0,L/a)$ at fixed scale $L=1.436\,r_0$
\label{TAB_zp}
}
\end{table}

%% file: zp60_tab.tex
\begin{table}[tb]
\centering
\begin{tabular}{r c c c}
\hline \\[-1.0ex]
      &         & $\ct^{\rm 1-loop}$ & $\ct^{\rm 2-loop}$  \\
$L/a$ & $L/r_0$ & $\zp(g_0=1,L/a)$ & $\zp(g_0=1,L/a)$ 
 \\[1.0ex]
\hline \\[-1.0ex]
 6 & 1.1177(46) &     0.5728(26)     & 0.5712(27) \\
 8 & 1.4903(61) &     0.5197(24)     & 0.5171(26) \\
10 & 1.8629(76) &     0.4590(31)     & 0.4584(30)
  \\[1.0ex]
\hline
\end{tabular}
\caption[TAB_zp60]{\footnotesize
Values of $\zp$ at $\beta=6.0$ for $L$ around
$L=1.436\,r_0$ 
\label{TAB_zp60}
}
\end{table}